\documentclass{aa}

\makeatletter

\renewcommand*\aa@pageof{, page \thepage{} of \pageref*{LastPage}}

\usepackage{natbib}
\bibpunct{(}{)}{;}{a}{}{,}%
\def\bibfont{\aa@bibliographyfont}%
\providecommand{\@LN}[2]{}

\makeatother

\usepackage{graphicx}
\usepackage{newtxtext}
\usepackage[smallerops,cmbraces]{newtxmath}
\usepackage{xcolor}
\definecolor{xlinkcolor}{cmyk}{1,1,0,0}
\usepackage[
  bookmarks=true,         
  pdfnewwindow=true,      
  colorlinks=true,        
  linkcolor=xlinkcolor,   
  citecolor=xlinkcolor,   
  filecolor=xlinkcolor,   
  urlcolor=xlinkcolor,    
  final=true,
]{hyperref}

\usepackage{bm} 
\usepackage{textgreek} 

\usepackage[capitalize]{cleveref} 
\crefname{section}{Sect.}{Sects.}
\creflabelformat{equation}{#2#1#3} 
\crefname{enumi}{item}{items} 

\usepackage{siunitx} 
\DeclareSIUnit[number-unit-product = ]\percent{\char`\%} 

\usepackage{csquotes} 

\usepackage{multirow}
\usepackage{siunitx,  
            booktabs, 
            caption}  
\usepackage{pifont}

\usepackage[caption=false]{subfig}
\definecolor{blackberry}{HTML}{8D1D75}

\DeclareSIUnit\parsec{pc}
\DeclareSIUnit\dex{dex}
\DeclareSIUnit\h{\mathnormal{h}}
\DeclareSIUnit\year{yr}
\DeclareSIUnit\years{yrs}
\DeclareSIUnit\arcsec{arcsec}
\DeclareSIUnit\arcmin{arcmin}
\DeclareSIUnit\Msun{M_\odot}
\DeclareSIUnit\Rsun{R_\odot}
\DeclareSIUnit\Lsun{L_\odot}
\DeclareSIUnit\Rvir{\mathnormal{R}_\mathrm{vir}}
\DeclareSIUnit\Rhalf{\mathnormal{R}_{1/2}}
\DeclareSIUnit\erg{erg}
\DeclareSIUnit\angstrom{\text{Å}}

\newcommand*{\Msun}{\ensuremath{\mathrm{M}_\odot}} 
\newcommand*{\Rsun}{\ensuremath{\mathrm{R}_\odot}} 
\newcommand*{\Lsun}{\ensuremath{\mathrm{L}_\odot}} 
\newcommand*{\Rvir}{\ensuremath{R_\mathrm{vir}}} 
\newcommand*{\Rhalf}{\ensuremath{R_{1/2}}} 

\newcommand*{\eshort}[2]{\ensuremath{#1 \cdot 10^{#2}}}

\defcitealias{ivleva+24}{\mbox{Paper I}}

\begin{document}

\title{Merge and Strip II: Imprint of galaxy formation physics and viscosity on baryon-dominated dwarf galaxies}
\titlerunning{Merge and Strip II: viscosity shaping dwarf galaxies in clusters}


\author{
    Anna Ivleva\inst{\ref{inst:usm}}
    \and
    Klaus Dolag\inst{\ref{inst:usm},\ref{inst:mpa}}
    \and
    Rhea-Silvia Remus\inst{\ref{inst:usm}}
    \and
    Duncan A. Forbes\inst{\ref{inst:swinburne}}
    \and
    Tirso Marin-Gilabert\inst{\ref{inst:usm},\ref{inst:cfa}}
}
\authorrunning{A. Ivleva et al.}

\institute{
    Universitäts-Sternwarte, Fakultät für Physik, Ludwig-Maximilians-Universität München, Scheinerstr. 1, 81679 München, Germany\label{inst:usm}\\
    \email{ivleva@usm.lmu.de}
    \and
    Max-Planck-Institut für Astrophysik, Karl-Scharzschild-Str. 1, 85748 Garching, Germany\label{inst:mpa}
    \and
    Centre for Astrophysics \& Supercomputing, Swinburne University, Hawthorn, VIC 3122, Australia\label{inst:swinburne}
    \and
    Center for Astrophysics | Harvard \& Smithsonian, 60 Garden St. Cambridge, MA 02138, USA \label{inst:cfa}
}

\date{Received 01 May, 2026 / Accepted XX Month, 20XX}

\abstract
{}
{}
{}
{Motivated by the discovery of peculiar dwarf galaxies inside galaxy clusters such as blue candidates (BCs), dark galaxies and ultra-diffuse galaxies (UDGs), we present hydrodynamic simulations of galaxy mergers in cluster environments. We vary the viscosity and stellar feedback prescriptions, realistically modelling possible conditions for hydrodynamic drag and fluid instabilities, as well as internal destabilization through stellar feedback-driven heating and gas loss. We find that long-lived tidal dwarf galaxies (TDGs) can form throughout all viscosity values applicable to galaxy clusters if stellar feedback is moderate. Our results expand on studies of cloud crushing simulations, investigating the entrainment problem in intracluster medium ambience. The smallest clouds have gas masses on the order of $M_{\rm gas}\sim10^7\,\Msun$ and reach relatively low final drift velocities of $\sim100\,\rm km/s$. The lowest possible Reynolds number acting on this class of clouds is $Re\sim1$ for full Spitzer viscosity. Almost all TDGs display elevated star formation rates of $0.01-0.1\,\Msun/\rm yr$, which are stable across several Gyr. Based on their matching properties, we support that BCs observed in the Virgo cluster are likely stripped TDGs. Similar features are also found in comparison with dark galaxies and baryon-dominated UDGs, implying that a subsample of these objects are also long-lived TDGs. This work provides robust evidence that stripping from galaxy mergers is a viable channel for the formation of stable cold gas clouds and dark matter-deficient galaxies observed in galaxy clusters.
}
{}


\maketitle
%

\section{Introduction}
\label{sec:introduction}

The number of peculiar dwarf galaxies with high baryonic mass fractions observed in the local Universe is steadily rising. Such objects have been found across all kinds of environments, appearing as completely isolated dwarfs \citep[e.g.,][]{Guo:2020}, as spheroidals nearby a larger galaxy \citep[e.g.,][]{Hammer:2020} and inside massive galaxy clusters. In particular, a subsample of ultra-diffuse galaxies (UDGs), which are characterized by an usually low surface brightness for their stellar size \citep[e.g.,][]{vanDokkum:2015,Koda:2015,Mihos:2015,Roman:2017,Forbes:2023,Gannon:2024,Zoeller:2024}, has been found to host significant HI content, while also displaying lower star formation efficiencies than other dwarf galaxies with similar gas masses \citep[e.g.,][]{Leisman:2017,Jones:2018,Kado-Fong:2022}. In some cases, high-resolution spectroscopic measurements allowed an analysis of their underlying gravitational potentials, which surprisingly revealed cases of UDGs that are dark matter-deficient or even entirely baryon dominated \citep[e.g.,][]{vanDokkum:2018,Toloba:2018,vanDokkum:2019,ManceraPina:2019,Buzzo:2025}. We refer to \citet{Gannon:2026} for a comprehensive overview of observed UDGs and their varying properties.

Another class of odd objects has been observed inside the Virgo cluster, where its vicinity enabled the detection of very small star forming dwarf galaxies with stellar masses around $10^4 - 10^5\,\Msun$, appropriately called blue candidates (BCs) or \enquote{blue blobs}. They have distinct clumpy morphologies, are metal-rich and have elevated star formation rates compared to other dwarf galaxies in their stellar mass range \citep[e.g.,][]{Jones:2022,Jones:2022b,Bellazzini:2022,Dey:2025}. Curiously, it has been found that BCs can have quite large HI reservoirs, which became apparent from two cases of dark galaxies, that turned out to actually host a small stellar component and were successively reclassified as BCs \citep{Adams:2015,Bellazzini:2015,Cannon:2015,Sand:2017,Junais:2021}. Dark galaxy candidates, i.e. dwarfs found exclusively though the \SI{21}{\centi\meter} line without a counterpart in optical or UV, lately started to be found in wide radio surveys, though the whole sky has not been entirely mapped yet \citep[e.g.,][]{Minchin:2005,Bilek:2020,Leisman:2021,Wong:2021,Jozsa:2022,Xu:2023,OBeirne:2024,Kwon:2025,Monaci:2026}. It is an unsolved question what the origin of these HI-dominated galaxies is, and why they show so little evidence for star formation.

Tidal dwarf galaxies (TDGs), i.e. dwarfs that are formed in the tidal tails of a galaxy merger event have been proven to display small dark matter fractions both in observations \citep{DucMirabel:1998,Rakhi:2023,Gray:2023} and simulations \citep{Barnes:1992,Ploeckinger:2018,Haslbauer:2019}. We have recently investigated the evolution of TDGs inside cluster environments through a series of simulations \citep[][from now referred to as \citetalias{ivleva+24}]{ivleva+24} and found that the drag induced by the intracluster medium (ICM) is causing significantly more TDGs per merger event. This leads to a lower mass limit for long-lived TDGs than predicted in the pioneering study by \citet{BournaudDuc:2006}, which did not include environmental influences. The resulting dwarfs then become independent by stripping away from the tidal tail, thus contributing to the dwarf population inside galaxy clusters. However, this result may change depending on the properties of the ICM, in particular on its ability to mix with the cold gas from the galaxy. This aspect is very sensitive to the viscosity treatment in the simulation code, which defines the range of Reynolds numbers one can resolve. So far, it is still unclear which range of Reynolds numbers $Re$ is applicable for galaxy clusters, which makes it difficult to infer their viscosity value. X-ray measurements of hot bubbles in static ICM imply values between $Re=62-400$, depending on the temperature one assumes for the medium \citep{Robinson:2004,Reynolds:2005}, suggesting that the system is turbulent. However, these flows are characterized by length scales that are several orders of magnitude larger than in the case of dwarf-sized gas clouds moving through a galaxy cluster. Their dynamics might therefore be characterized by much smaller Reynolds numbers, allowing them to persevere through the lack of violent turbulence.

In fact, extensive efforts have been made during the past 20 years in order to constrain the criteria that define stable configurations between fluids, in the effort to explain a number of different observed astrophysical phenomena involving multiphase gas. Early works mostly focused on the behavior of hot, AGN-inflated bubbles in the ICM, finding that already a fraction of the Spitzer value as the effective viscosity coefficient can quench Rayleigh-Taylor and Kelvin-Helmholtz instabilities, while magnetic fields can additionally stabilize the fluid interface \citep{BrueggenKaiser:2001,DeYoung:2003,Reynolds:2005,SijackiSpringel:2006,Roediger:2007}. A similar effect was found in the confinement efficiency of cold, ram pressure stripped gas clumps in the wake of jellyfish galaxies, which can develop long, star forming filaments that stay connected to the disk galaxy while moving through an ICM \citep[e.g.,][]{Ruszkowski:2014,Tonnesen:2014,Roediger:2015,Kraft:2017,Sparre:2024,Marin-Gilabert:2024}. Later on, multiphase mixing between separate cold or hot clouds and hot ambient medium became an active field of research, driven by the discovery of outflows in galaxies and high velocity clouds around the Milky Way \citep{Wakker:1991,Thom:2008}. Often referred to as \enquote{cloud-crushing} simulations, these studies revealed the immense complexity of the entrainment problem, to which still no consistent solution has been found explaining the observed stability of gas clouds \citep[e.g.,][]{GronkeOh:2018,Li:2020,Kanjilal:2021}. Also here magnetic fields were found to stall destructive turbulence \citep{Hidalgo-Pineda:2024,Kaul:2025,Sparre:2020}. Interestingly, the outcome may also differ when considering a loosely associated group instead of a single cloud, which can lead to stabilization through multi-cloud shielding \citep{Melioli:2005,ForbesLin:2019,Aluzas:2012}. As pointed out in a recent review by \citet{GronkeSchneider:2026}, these works mostly focus on the problem of outflowing clouds, while infalling scenarios have been studied in far less detail, though such circumstances may lead to more destructive conditions \citep{Heitsch:2009,Heitsch:2022,Afruni:2019,Gronnow:2022,Tan:2023}. In addition, these simulations primarily focused on environments applicable to the interstellar and circumgalactic medium. The behavior in ICM-like atmospheres such as in this work is still widely unexplored.

In this paper, we investigate the evolution and stability conditions for cold gas clouds and baryon-dominated dwarf galaxies inside galaxy clusters by running a series of targeted, high-resolution radiative hydrodynamic simulations. The paper is structured as follows: we abridge the setup in \cref{sec:setup} and outline the relevant subgrid modules used in the simulation. In \cref{sec:results} we present our results, focussing on the general mixing trends across varying physical realizations, before examining the emerging dwarf galaxy populations and their evolution in mass, star formation rate and appearance. We discuss the consequences of our findings by comparing to recent results in cloud crushing studies and finally by proposing analogs to observed dwarf galaxies. We summarize our conclusions in \cref{sec:sumcon}.

\section{Numerical setup} \label{sec:setup}

We initialize a major merger with a mass ratio of \mbox{2:1} between two late-type galaxies inside a galaxy cluster, where the more massive galaxy has a Milky Way-like mass of $M_{200,\rm c}=10^{12}\,\Msun$. The bulk motion of the galaxies is set to correspond to an elliptical orbit inside the cluster, where their initial drift velocity is taken to be the cluster's virial velocity ($=670\,\rm km/s$, corresponding to a virial mass of $M_{200,\rm c}=10^{14}\,\Msun$). The initial distance of the galaxies to the cluster center is two times its virial radius, such that a galaxy encounter is still realistic due to relatively low dispersion velocities outside the galaxy cluster. The cluster is modelled by dark matter and ICM particles, where the latter are in hydrostatic equilibrium, while their resolution is $m_{\rm dm}=\eshort{3.2}{6}\,\Msun$ and $m_{\rm dm}=\eshort{6.6}{5}\,\Msun$ for dark matter and baryons, respectively. The initial conditions are identical to simulation C45 presented in \citetalias{ivleva+24} and we refer to its Section 2 for the detailed description of the setup.

\subsection{Baseline subgrid physics}

The simulations were carried out with the cosmological code \mbox{\textsc{OpenGadget3}}, which utilizes a smoothed particle hydrodynamics (SPH) scheme to model gas dynamics and a Tree solver for gravity (release paper by Dolag et al. in prep.). Its infrastructure is based on its predecessor Gadget2 \citep{Springel:2005} but employs major improvements in the gravity solver \citep{Ragagnin:2016} and SPH implementation \citep{Beck:2016,Pakmor:2012}. We used the Wendland C4 function as the kernel to reconstruct the local hydrodynamic properties of the simulated fluid based on the position and smoothing length of the simulated particles \citep{Wendland:1995,Dehnen:2012,Donnert:2013}. We include spatially varying conduction modeling heat exchange \citep{Beck:2016}, and second-order terms in velocity gradient estimations, which significantly improves convergence between SPH and grid codes in turbulent flows. The code takes into account cooling for an optically thin, primordial composition gas in ionization equilibrium with a UV background \citep{Katz:1996,Springel:2005} and a multiphase model for star formation based on the effective equation of state for the star forming gas particles \citep{SpringelHernquist:2003}.

\subsection{Varied subgrid physics}

Even when modelling an ideal fluid, SPH implementations usually include an additional term in the Euler equations, which can be interpreted as bulk viscosity of the medium and is commonly referred to as artificial viscosity \citep{Balsara_1995, Cullen_2010}. Its purpose is to stabilize the code numerically and properly capture the behavior at shock interfaces, which otherwise would be incorrect due to the entropy-conserving formulation of SPH. Outside of shocks, the additional term should decay to zero when modelling an ideal fluid. We test two realizations of artificial viscosity in this work, which we refer to as \enquote{classical} and \enquote{improved} SPH. Both versions rely on a built-in shock finding scheme, which boosts the local bulk viscosity coefficient for an interacting SPH particle pair when a shock is detected. However, it has been established that the effective bulk viscosity in \enquote{classical} SPH tends to be non-vanishing even outside shock regions, effectively modelling a viscid fluid. \enquote{Improved} SPH therefore includes an additional time-depending  moderator in the artificial viscosity term, which efficiently decays the effective viscosity after a shock has passed, returning to an ideal fluid representation. This way, low Reynolds numbers are reached only in shock regions, while high Reynolds numbers are resolved elsewhere \citep{Price:2012}.

In addition we also test the effect of physically motivated viscosity, i.e. when the fluid is in fact non-ideal and represented by the Navier-Stokes equations, including non-vanishing shear viscosity \citep{Marin-Gilabert:2022}. Additional artificial viscosity is still necessary in this regime for accurate shocks capturing via bulk viscosity \citep{SijackiSpringel:2006}. We use full Spitzer viscosity as the maximum shear viscosity coefficient \footnote{$\eta=\eshort{6}{-17}T[K]^{5/2}\frac{\rm g}{\rm cm\,s}$, assuming a Coulomb logarithm of \mbox{$\ln\Lambda=37$} \citep{Spitzer:1965}.}.

Two feedback modes are included by the star formation model used in our simulations. A star forming gas particle has an associated cold and hot phase, which can transfer mass and energy among each other \citep[for a detailled description, see][]{SpringelHernquist:2003}. In particular, it includes a heating rate motivated by canonical supernova release energies, which can also be interpreted as isotropic \enquote{thermal} feedback on the surrounding gas via conduction. Additionally, it is possible to switch on \enquote{kinetic} feedback, which in its original implementation is supposed to capture galactic winds, thus accounting for starbursts and associated outflows observed in galaxies. It should be noted, however, that the parameters associated with this subgrid module have been gauged for galaxies that are at least in the Milky Way mass regime, not for dwarf galaxies. Hence, the effective galactic wind modelled by the \enquote{kinetic} feedback mode should be interpreted as the result of extreme starburst episodes that may occur in a dwarf galaxy, while the \enquote{thermal} mode is representative of more moderate stellar feedback.

\section{Results}\label{sec:results}

We implement four different realizations in total, which cover the feasible parameter space of both the (i) mixing behavior of multiphase gas in cluster atmospheres by varying the viscosity prescription and (ii) feedback induced gas heating and ejection through different stellar feedback mechanisms. \cref{tab:simtable} summarizes the differences between the simulations, where the meaning of each column was introduced in \cref{sec:setup}. Simulations A and D represent a very viscous ICM, which they achieve by setting a high bulk viscosity (A, numerically motivated) or high shear viscosity, given by the full Spitzer value (D, physically motivated). Opposite to that are setups B and C, which model a practically inviscid ICM, where simulation C resolves additional kinetic stellar feedback. Thermal feedback is included in all four setups.

\begin{figure*}[h!]
    \centerline{\includegraphics[width=0.85\textwidth, trim={0 0 0 0}]{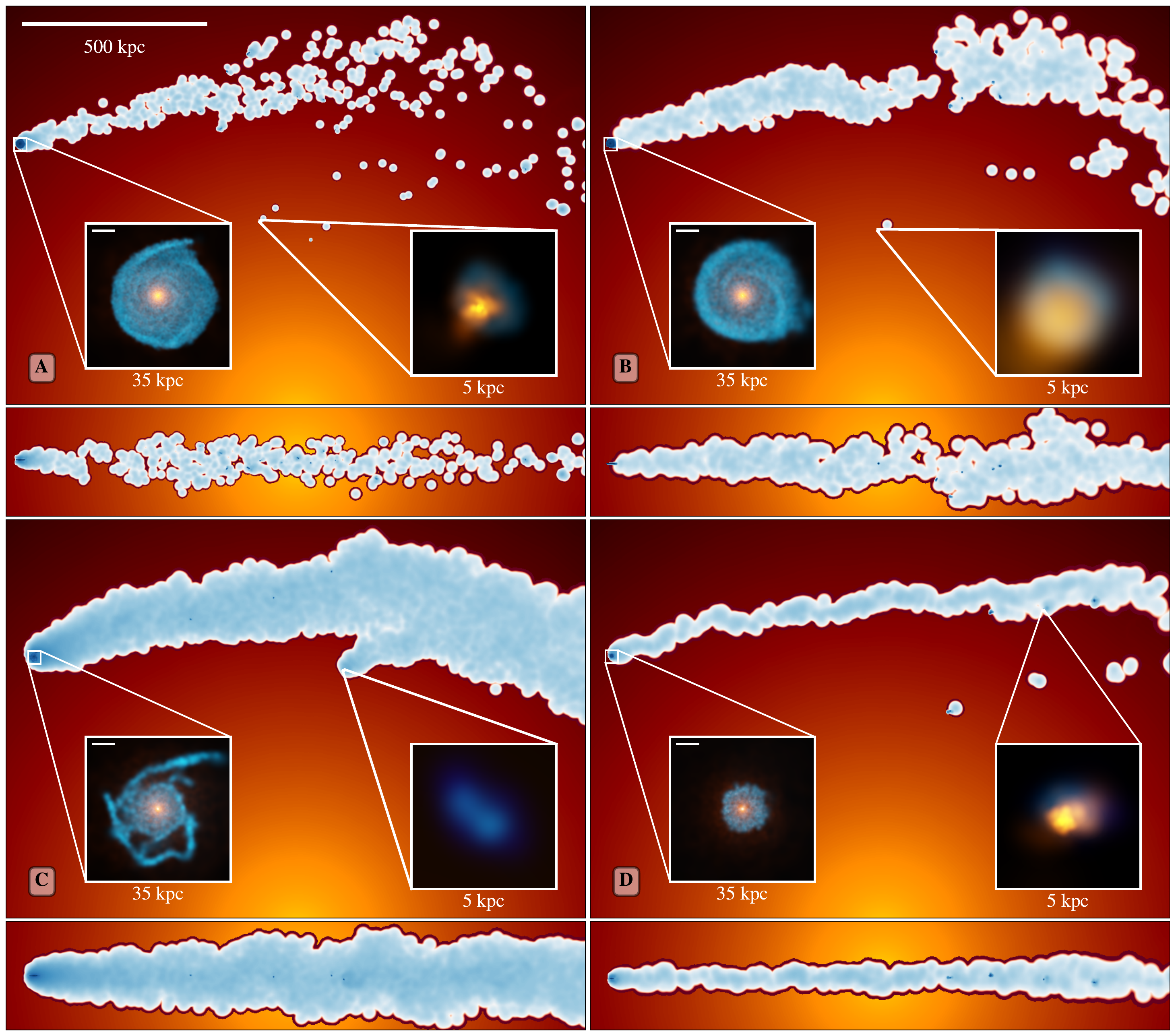}}
    \caption{Qualitative comparison of ram pressure stripped gas in different subgrid physics in face-on and edge-on projection (top and bottom panel of each quadrant, respectively). The letter in the bottom left of each quadrant indicates the respective simulation (see \cref{tab:simtable}). The red-yellow background represents the surface density of the ICM, while light-blue traces the gas stripped from the merging galaxies. Two inset images in each simulation enhance on the merger remnant and an example dwarf galaxy. The scale in the top left of each merger inset image is indicating a size of \SI{5}{\kilo\parsec}.}
    \label{fig:zoomoutcomp}
\end{figure*}

\subsection{General gas behavior}\label{subsec:generalgasbehavior}

\cref{fig:zoomoutcomp} shows a comparison of all four simulations, which illustrates the qualitative differences in the evolution of the merger remnant, ram pressure-stripped gas, and tidal dwarf galaxies (TDGs). At the moment of this figure, the coalescence event was $\sim$\SI{1.3}{\giga\year} ago. In each case we display both the face-on and edge-on view of the merger orbit in the top and bottom panel of each quadrant, respectively. The background in red and yellow is indicating the integrated surface density of the intracluster medium (ICM) along the line of sight, while the blue areas represent ram pressure-stripped gas from the merging galaxies. In order to visualize the qualitative behavior, we determine the surface density of the ram pressure-stripped component separately from the ICM, since the density contrast would otherwise be too low to distinguish them. This is the reason for the spot-like appearance in places where the stripped gas density is very low, encompassing the size of the smoothing kernel of the respective SPH particle. The two inset images in each simulation display a zoom-in on the merger remnant (left) and an example dwarf galaxy (right), where the side lengths of the images are \SI{35}{\kilo\parsec} and \SI{5}{\kilo\parsec}, respectively. For reference, we plot a scale in the top left corner of each merger zoom-in image representing \SI{5}{\kilo\parsec}, while the scale in the top left of the figure is indicating a size of \SI{500}{\kilo\parsec} ($=0.5 R_{\rm 200, c}$, where $R_{\rm 200, c}$ is the galaxy cluster's virial radius). The zoom-in images were created with the ray-tracing visualization software \textsc{Splotch} \citep{Dolag:2008} and in this case render gas in blue, while stars are shown in orange.

\begin{table}[t!]
\caption{Overview of the implemented subgrid physics in the different runs compared in this work. Crosses mark a switch chosen between mutually excluding options per category, while checks are placed where the respective subgrid physics is optional.} \label{tab:simtable}
\centering
\begin{tabular}{@{} l *{5}{c} @{}}
\toprule
& \multicolumn{2}{c}{stellar feedback}
& \multicolumn{2}{c}{artificial viscosity}
& \multirow{2}{*}{Physical}\\
\cmidrule(lr){2-3} \cmidrule(lr){4-5}
Sim & thermal & kinetic & classical & improved & viscosity\\ [2mm]
 \toprule
$\rm A$ & \ding{51} &  & \ding{55} &  &  \\
$\rm B$ &  \ding{51}  &  &  & \ding{55} &  \\
$\rm C$ & \ding{51} & \ding{51}&  &  \ding{55} &  \\
$\rm D$ & \ding{51} &  &  & \ding{55} & \ding{51} \\
\bottomrule
\end{tabular}
\end{table}

While the viscous realizations (A and D) display a relatively narrow ram pressure-stripped gas tail, B and C exhibit wider trails -- already indicating that mixing between the cold galactic gas and hot ICM is occurring on faster timescales there. This effect is further enhanced when kinetic feedback is switched on in simulation C, where efficient stellar winds are acting as an additional channel to expel gas from the galactic disk apart from ram pressure. A look at the zoom-in of the merger remnant confirms that the gaseous disk is still struggling to find an equilibrium after the merger due to the strong feedback, while the disks in the other simulations have already settled. However, while A and B maintained a relatively large gas reservoir, the remnant in simulation D is left with a much smaller disk with about half the radius. This is caused by the interplay of two circumstances: (i) setup D includes an explicit shear viscosity coefficient that rises with sharp velocity gradients, which supports gas stripping through Kelvin-Helmholtz instabilities. The formulation of the numerical viscosity in setups A-C, on the other hand, is coupled to velocity divergence, representative of bulk viscosity. (ii) The disk edge displays high viscosity coefficients in simulation D, driven by the strong temperature dependence on the Spitzer value and kernel averaging over cold disk and hot ICM particles. Meanwhile, the appearance of the dwarf galaxy is remarkably similar when comparing A and D, whereas the dwarf's gas component in the inviscid simulations (B and C) is more diffuse, leading to less effective star formation and hence lower surface brightness.

\subsection{Shock and mixing behavior in the ICM}\label{subsec:excess}

\begin{figure*}[ht!]
    \includegraphics[width=0.49\textwidth, trim={0 0 0 0}]{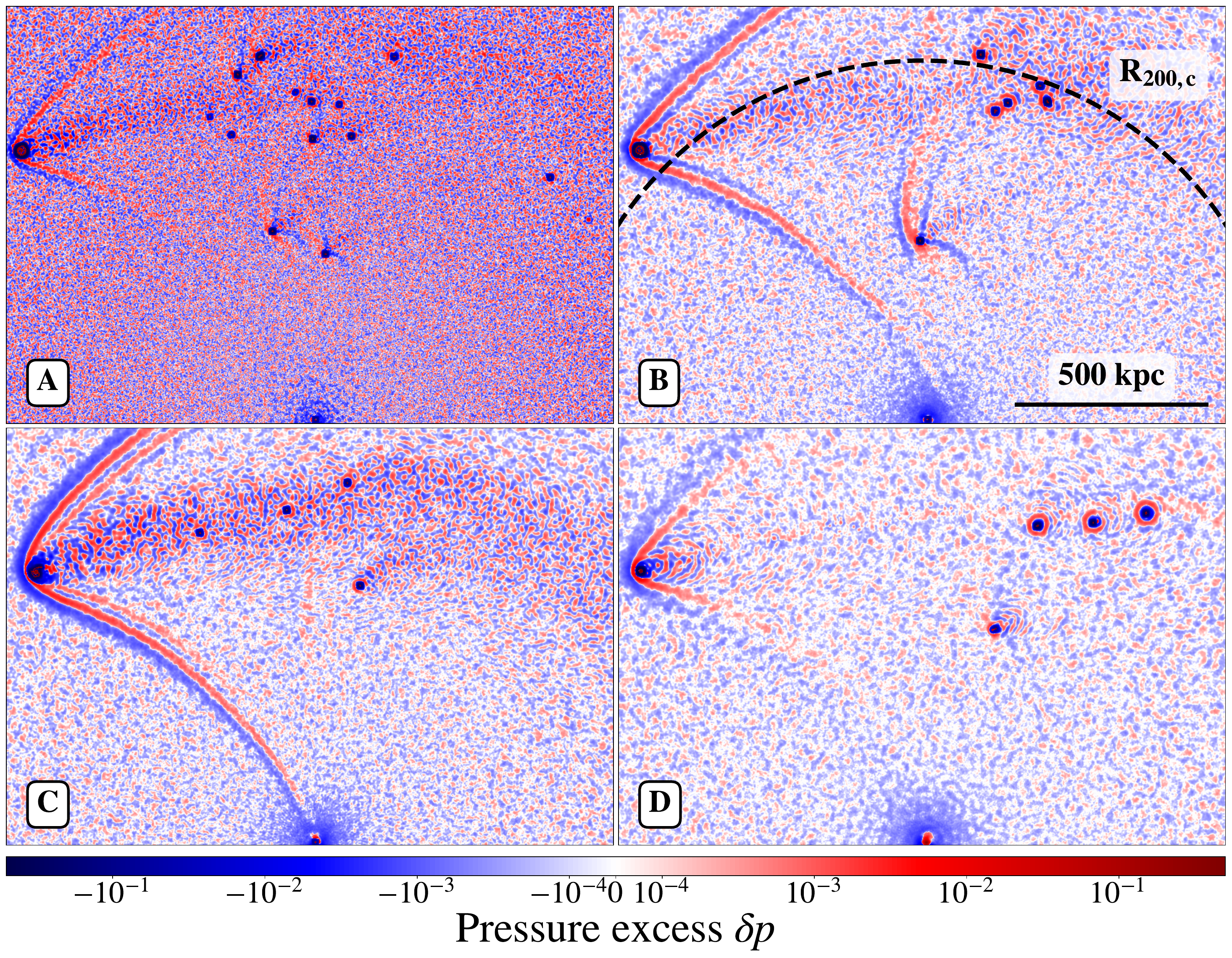}
    \includegraphics[width=0.49\textwidth, trim={0 0 0 0}]{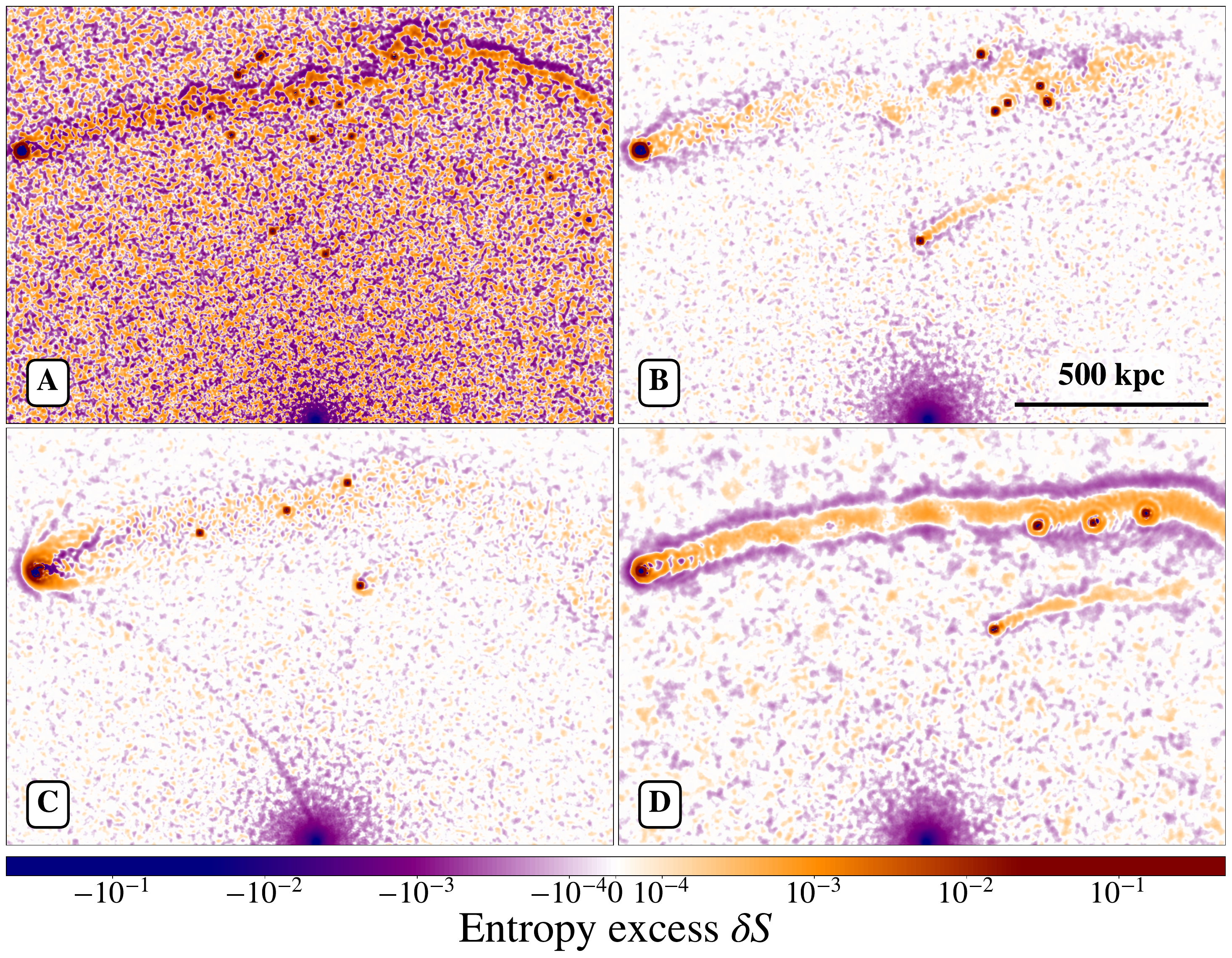}
    \caption{Pressure (left) and entropy (right) excess for all four simulations analyzed in this work. The letter in the bottom left corner of each panel indicates the setup (see \cref{tab:simtable}). The scale in panel B denotes \SI{500}{\kilo\parsec}, while the dashed line indicates the virial radius $R_{\rm 200,c}$ of the galaxy cluster. The merger remnant as well as the most massive and fastest dwarf galaxies drive a pressure wave through the cluster. The ram pressure stripped gas tails in simulations A and D (viscous) are relatively stable, while mixing in B and C (inviscid) quickly destroys them.}
    \label{fig:excess}
\end{figure*}

Different subgrid implementations, intended to capture different astrophysical circumstances, may lead to varying properties and dynamics across different scales. Before investigating these differences in the evolution of dwarf galaxies formed in our simulations in the following sections, we will focus first on the consequences on the ICM. In particular, we will investigate the buildup and strength of the various pressure waves triggered by the galaxies moving through the cluster, as well as the emerging mixing behavior between the cold gas phase stripped from the galaxies and the hot surrounding medium. Particularly the mixing regime and hence the timescales at which turbulence can disrupt structures will heavily affect the appearance and evolution of gaseous dwarf galaxies.

In order to pinpoint these features, we need to find the position of pressure and entropy jumps in the ICM, which indicates the interfaces of shocks and mixing layers, respectively. For this purpose, we use unsharp masking of property maps, as this technique has proven now to be an effective tool to visualize discontinuities \citep[e.g.,][]{Fabian:2003,Dolag:2005,SijackiSpringel:2006,Sommer:2024,Ivleva:2026}. The process is the following: starting from a property map $\mathcal{X}(x,y)$, we smooth the image by applying a 2D Gaussian kernel function $G(x,y)\propto \exp[-(x^2+y^2)/(2\sigma^2)]$ to each pixel of the original image, which yields a blurred version $\tilde{\mathcal{X}}(x,y)$. By subtracting the original map from the smoothed variant, sharp value jumps remain, while features that are larger than the smoothing kernel $\sigma$ will cancel out. Note that this will only work if the features one is looking for are characterized by a sufficiently large value contrast with respect to the background, as the smoothing kernel will attenuate these jumps. Lastly, the contribution of each pixel is normalized by the local value of the blurred image, simply to ensure that all edge features appear equally strong in the resulting map, no matter the contrast. The final discontinuity map of a property $\delta \mathcal{X}$ is therefore given by $\delta \mathcal{X} = (\mathcal{X}-\tilde{\mathcal{X}})/\tilde{\mathcal{X}}$, where $\mathcal{X}$ will be pressure $p$ or entropy $S$, tracing shock and distinct mixing layers, respectively. Local pressure is evaluated by $p\propto\rho \cdot T$, while entropy is taken to be $S\propto T / \rho^{2/3}$, where $\rho$ and $T$ are the density and temperature integrated along the line of sight, respectively. Due to the normalization of $\delta \mathcal{X}$, simply the proportionality on density and temperature is already sufficient for creating the excess maps.

\cref{fig:excess} presents the pressure and entropy excess map for each simulation on the left and right side, respectively, where the letter in the bottom left corner of each panel is specifying the simulation (see \cref{tab:simtable}). The time is the same as in the overview in \cref{fig:zoomoutcomp} (\SI{1.3}{\giga\year} after merger event), while the scale in the bottom right of configuration B is indicating a length of \SI{500}{\kilo\parsec} for each simulation. For reference, the dashed line is marking the virial radius $R_{\rm 200, c}$ of the galaxy cluster, of which the center is apparent in each panel as the dark spot in the bottom of each panel. The system is shown in its face-on projection, i.e. the merger orbit is lying in the sky plane. We used a Gaussian smoothing kernel with a size of $\sigma=5$ pixel in these maps.

\subsubsection{Pressure excess $\delta p$ tracing sound waves}

The merger remnant, clear as the strong disturbance near the left border of each panel, has already undergone its pericenter passage at this stage and has just moved outside the cluster boundary of $R_{200,c}$. The cone around the remnant is a consequence of its relatively high velocity with respect to the ICM. It is leading to the buildup of a pressure (sound) wave at the front of the galaxy, which is then further propagating through the whole cluster. The red and blue pattern of the cone is evident for this, signifying a depression (blue) vs. excess (red) of pressure compared to the background, which is the ICM in hydrostatic equilibrium. We note that, although this looks like a shock wave, the pressure jump is not large enough to qualify as a shock.

As a matter of fact, the merger remnant is not the only object here, which is able to leave such a footprint in the ICM. Some of the stripped tidal dwarf galaxies, which are moving in the wake of the merger remnant and are clearly visible in the pressure excess map, are actually also accompanied by a cone, as well as an additional compression at their front. Note that it is not symmetric, but is elongated towards the cluster outskirts instead, because the dwarfs tend to move on a hook-like orbit, where they are quickly decelerated by ram pressure and continue to move on increasingly radial orbits. The reason why these dwarf galaxies display such a pressure wave is because they maintain the highest velocities with respect to the ICM, while also having large enough masses to disturb the medium. In fact, this is also correlated to their position in the cluster, as these dwarfs are closer to the cluster center in both simulations A and B. When they were stripped during the merger event, these dwarf galaxies had the largest gas mass out of their sample. They quickly lost their angular momentum while simultaneously converting a large portion of their gas reservoir into stars, leading to relatively radial orbits, but still having large velocities because they continued to move as stellar dominated objects and thus are not susceptible to hydrodynamic drag anymore.

Thus, it is an interesting observation that the dwarf galaxies in simulation D do not display such a pressure wave, though being in a similar mass range as the dwarfs with such a feature in simulation A and B (c.f. \cref{subsubsec:massevol}). It turns out, that although the intrinsic properties of these dwarf galaxies are similar, their dynamical features with respect to the cluster are different. While also representing a viscous fluid as in simulation A, the ICM in setup D slows down the dwarf galaxies on faster time scales. Figuratively speaking, star formation is not outrunning deceleration by ram pressure and turbulent drag anymore (discussed in more detail in \cref{subsec:cc_comp}), and thus the dwarfs' orbit in the cluster is affected longer by interaction with ICM. Their velocity with respect to the cluster is hence smaller than of the sample in A and B, leading to the lack of a pressure signature. The dwarfs in simulation C, on the other hand, actually have relatively high velocities. However, they have much smaller total masses than the galaxies in the other samples (A, B, D), and are accordingly not able to disturb the ICM.

\subsubsection{Entropy excess $\delta S$ tracing mixing}

Moving on towards the right side of \cref{fig:excess}, we now focus on the mixing behavior of the different gas phases inside the galaxy cluster, evident by the entropy excess map. Here, depression or excess with respect to the background indicate where the medium contains inhomogeneities, leading to locally varying entropy values across the different gas phases. In other words, orange and violet features will appear at locations where turbulence has not caused stripped gas to fully mix with the ICM yet.

When comparing the four different simulations, it immediately becomes clear that the merger remnant galaxies in setups A and D display the most pronounced ram pressure stripped gas tails. Particularly in simulation D it is quite persistent, stretching across the whole cluster. This is a clear signature of viscosity acting in these simulations, which is impeding turbulent mixing in simulation A, and even effectively confining the \enquote{jellyfish tail} in simulation D. In fact, this finding aligns well with the behavior of AGN-driven bubbles found by \citet{ScannapiecoBrueggen:2008}, who investigated the evolution of buoyant, underdense bubbles in an ICM. They found that the introduction of viscosity can actually impede the fragmentation of a cloud by creating turbulent mixing layers around the structure, which effectively confines the bubble during its rising motion.

In the meantime, the setups which realize low effective viscosities behave quite differently. Simulation B displays only very weak entropy excess features in the merger remnant trail, while simulation C lacks them entirely. Though these two simulations incorporate the same artificial viscosity subgrid model, this difference is caused by the additional kinematic stellar feedback in C, which aids mixing by expelling gas further out with respect to the line drawn by the merger orbit, leaving little opportunity for the gas tail to stabilize itself against turbulence.

Curiously, one can even see the tails of some dwarf galaxies, which is indicating their gas loss due to ram pressure stripping, in particular for the dwarfs towards the center of the panel in simulation B and D. Still, also here the slower mixing behavior in D is evident by the longer trail.

\subsection{Properties of simulated dwarf galaxies}\label{subsec:props}

Having analyzed the global behavior and dynamics of the emerging dwarf galaxy population inside the cluster, we will now focus on their internal properties. 
We will analyze their star formation rate (SFR) and mass evolution, before determining the dwarf galaxy type based on observational criteria.

\subsubsection{Star formation rate}\label{subsubsec:sfr}

\begin{figure}[t!]
    \centerline{\includegraphics[width=0.45\textwidth, trim={0 0 0 0}]{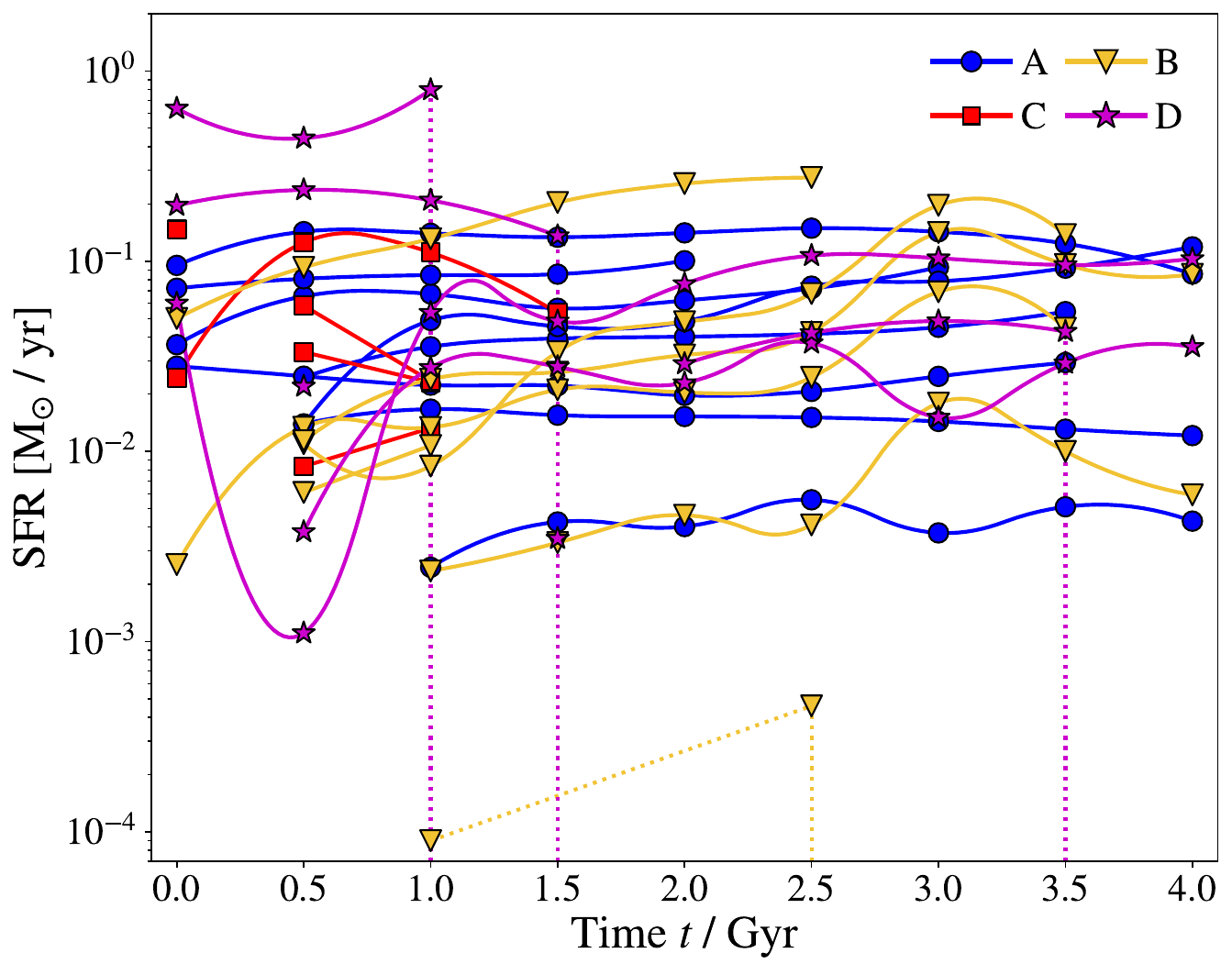}}
    \centerline{\includegraphics[width=0.45\textwidth, trim={0 0 0 0}]{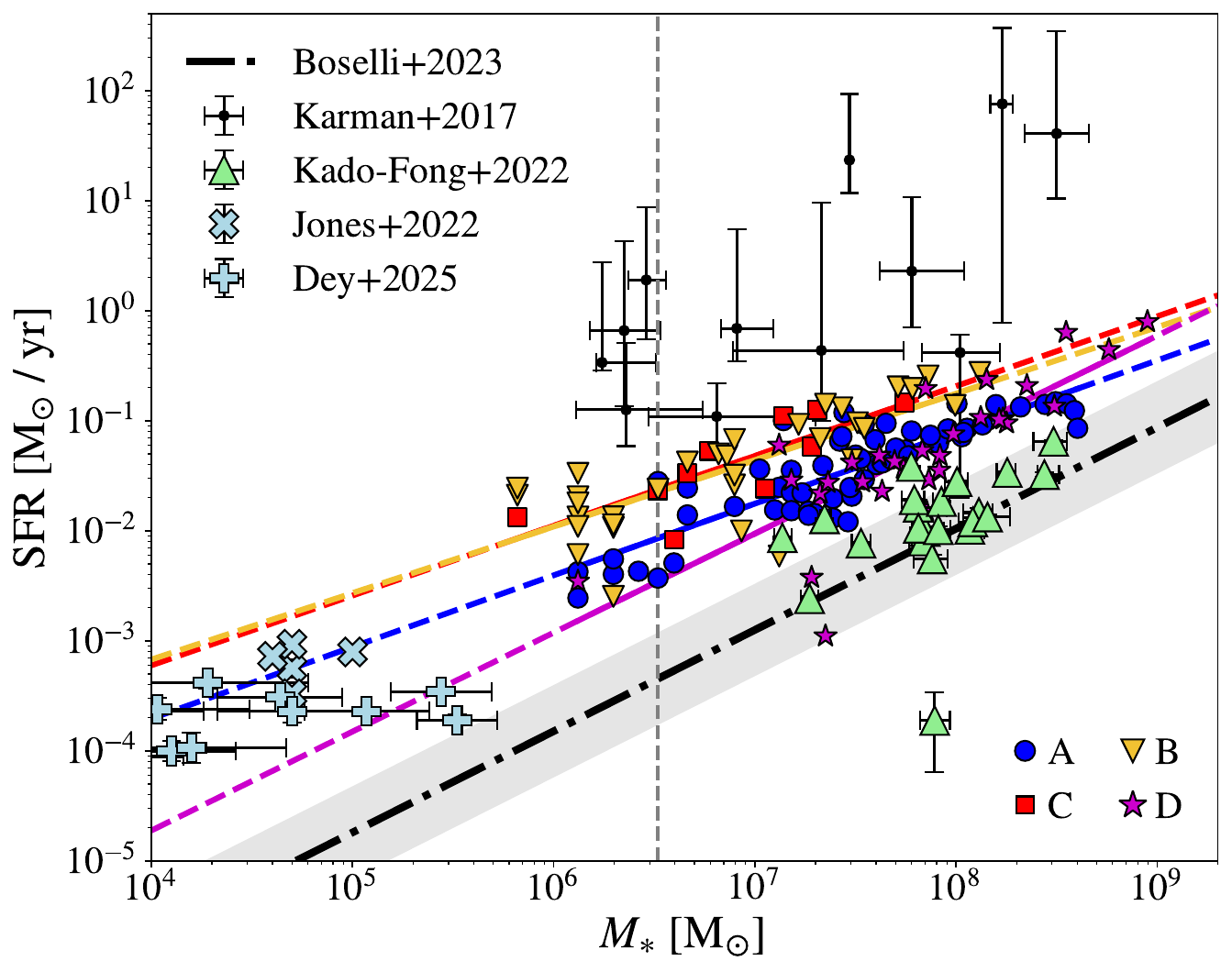}}
    \caption{Upper panel: Star formation rate SFR vs. time $t$ of simulated dwarf galaxies. Vertical dotted lines indicate events, where the dwarf was suddenly quenched. Lower panel: SFR vs. stellar mass $M_{\ast}$. The vertical dashed line indicates a threshold of five stellar particles in the simulation, while the four colored lines are fits to the dwarfs from the respective simulation. The black line shows the star-forming main sequence observed inside the Virgo cluster by \citet{Boselli:2023}. Black markers indicate dwarfs at high redshift by \citet{Karman:2017}, green triangles are HI rich UDGs by \citet{Kado-Fong:2022} and light-blue markers are \enquote{blue candidates} by \citet{Jones:2022} and \citet{Dey:2025}.}
    \label{fig:sfr}
\end{figure}

\begin{figure*}[ht!]
    \centerline{\includegraphics[width=0.9\textwidth, trim={0 0 0 0}]{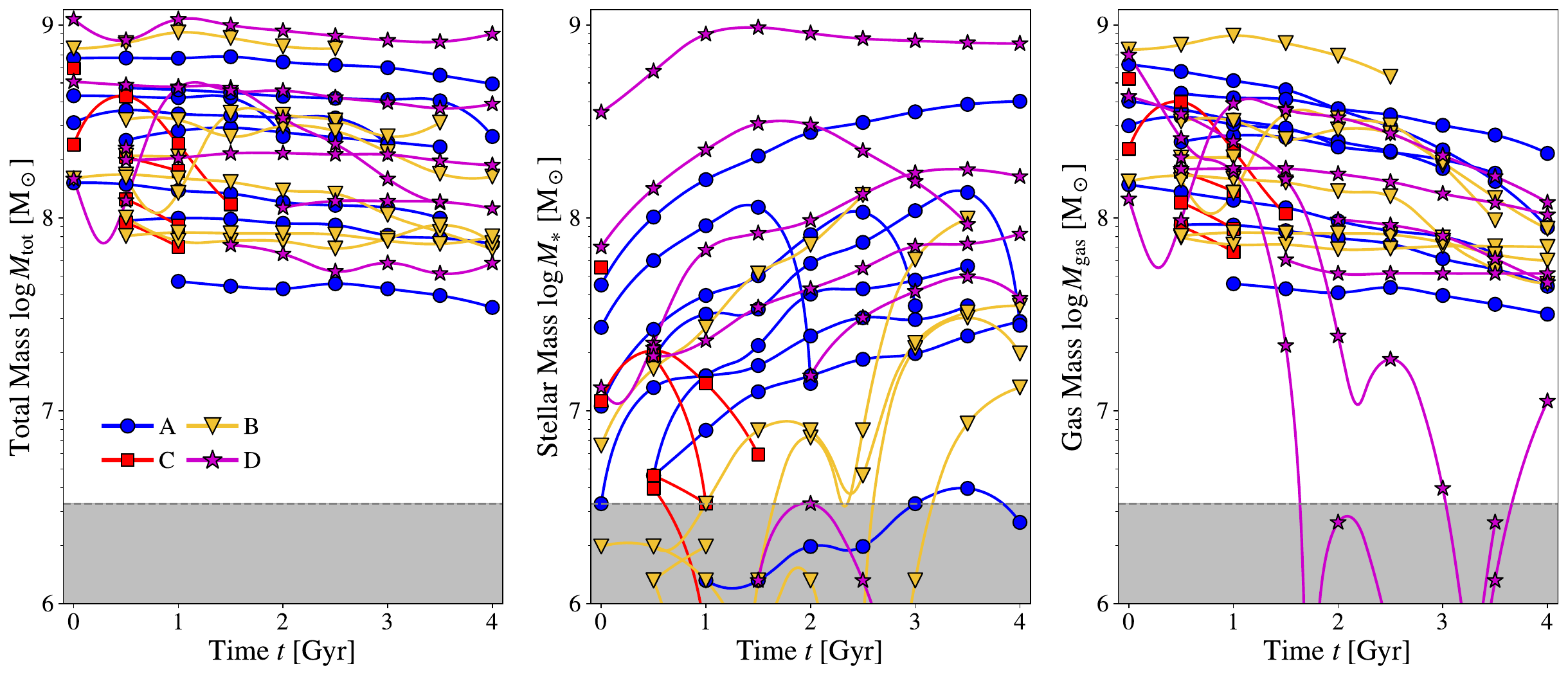}}
    \caption{Mass evolution for dwarf galaxies in the tested subgrid implementations over \SI{4}{\giga\year}. Left, middle and right panel show total mass $M_{\rm tot}$, stellar mass $M_\ast$ and gas mass $M_{\rm gas}$, respectively, while the horizontal dashed line indicates a threshold of five particles. The viscous simulations A and D produce dwarf galaxies with relatively constant total masses, where their gas content is continuously converted into stars. The objects in simulation B experience delayed star formation bursts after a few Gyr of existing as quiescent dark galaxies. The additional winds in simulation C efficiently destabilizes and destroys the dwarfs during the first \SI{2}{\giga\year}.}
    \label{fig:massevol}
\end{figure*}

\cref{fig:sfr} demonstrates the SFR of the simulated dwarf galaxies vs. time $t$ (upper panel) and stellar mass $M_{\ast}$ (lower panel). The four simulations are distinguished by different colors and symbols -- blue circles for A, yellow triangles for B, red squares for C and purple stars for D (see \cref{tab:simtable} for the differences in subgrid physics between the setups). The dwarfs' properties were traced over the course of \SI{4}{\giga\year} in steps of \SI{0.5}{\giga\year}, starting from the moment when the merger has just occurred and stripping of its tidal dwarf galaxies is beginning to take place. For reference, this is \SI{1.3}{\giga\year} before the moment when we analyzed the global gas behavior in \cref{subsec:generalgasbehavior,subsec:excess}. In order to make the figure more accessible, we fitted splines to the data points for smooth evolution lines in the upper panel. Therefore, the exact course of the lines between the markers is not necessarily representing the actual evolution, but rather is intended to guide the eye between the tracing times highlighted by the markers. The lower panel shows the location of the simulation dwarfs with respect to the star forming main sequence, taken at the same time steps as in the SFR evolution above. The vertical gray line is indicating a threshold of five particles contained by a simulated dwarf galaxy \mbox{($\eshort{3.3}{6} \,\Msun$)}, while the dashed-dotted black line with associated gray-shaded region is the main sequence for dwarf galaxies observed inside the Virgo cluster by \citet{Boselli:2023}. Additionally, we compare to several \enquote{unusual} dwarf galaxy types: the black markers represent a high redshift ($3<z<6$) star-forming dwarf survey by \citet{Karman:2017}, the light-green triangles are HI rich ultra-diffuse galaxies (UDGs) by \citet{Kado-Fong:2022} and the light-blue crosses and pluses are blue candidates (BCs) by \citet{Jones:2022} and \citet{Dey:2025}.

Overall, the simulated dwarfs display remarkably stable SFRs over the whole tracing period. Occasional cases of abrupt quenching (vertical dashed lines in upper panel of \cref{fig:sfr}) are either due to disruption, or to sudden stripping of the dwarf's gas component (c.f. \cref{subsubsec:massevol}). Setup B, however, displays rising SFRs over time, which is surprising considering the efficient mixing of the merger remnants gas tail (\cref{fig:excess}). But these dwarf galaxies are relatively stable and display delayed star formation with respect to the other simulations. Indeed, this aligns well with findings by \citet{Marin-Gilabert:2025}, who found that the local cooling efficiency is relatively constant despite varying viscosity coefficients, with a tendency towards more effective cooling in very viscous fluids. Moving on towards the bottom panel of \cref{fig:sfr}, it becomes apparent that the simulated dwarf galaxies display a systematically elevated SFR compared to the star forming main sequence observed in the Virgo cluster by about one order of magnitude. Meanwhile, the viscous setups (A and D) notably show a lower SFR for a given mass than in B and C. Interestingly, the slope of simulation D with physically motivated viscosity is very similar to the observed main sequence, while the runs with artificial viscosity only -- introduced to capture shocks in the simulation -- all have a shallower slope.

Despite having large gas reservoirs, \citet{Kado-Fong:2022} found that the UDGs in their sample have lower star formation efficiencies than other low-surface brightness galaxies with the same HI content. Still, they mostly lie on the main sequence, though some cases display elevated SFRs, becoming comparable to our simulated galaxies from setups A and D. Dwarfs at high redshift, on the other hand, display a significantly elevated SFR for a given stellar mass, similar to what has been already found for more massive galaxies. They lie above our simulated sample, though still exhibiting a small overlap at the low SFR end. Finally, we observe a remarkable agreement between our simulated sample and BCs, when extrapolating our results beyond our resolution limited lower mass end towards $\sim$$10^5\,\Msun$. In particular, the extensions of the viscous simulations A and D overlap nicely with these highly star forming clumps that are observed inside the Virgo cluster.

\subsubsection{Mass evolution} \label{subsubsec:massevol}

\cref{fig:massevol} presents the mass evolution of all dwarf galaxies in each of the four simulations, where the left, middle and right panel shows their total mass $M_{\rm tot}$, stellar mass $M_\ast$ and gas mass $M_{\rm gas}$, respectively. Note that the dwarfs are almost entirely baryon dominated, therefore their total mass can be approximated by $M_{\rm tot}\approx M_{\ast} + M_{\rm gas}$. The gray-shaded region in all panels is indicating a threshold of less than five particles contained by the dwarf galaxy in the simulation. As in \cref{fig:sfr}, we fitted splines to the data points, in order to better visualize their trend.

Interestingly, the overall evolution of their total mass is very similar for most dwarf galaxies across the simulations -- the only exception being setup C, where the objects are rapidly declining in mass within the first \SI{1.5}{\giga\year}. In contrast to that, the dwarf galaxies from the other three simulations (A, B, D) exhibit relatively constant halo masses for \SI{4}{\giga\year}, although displaying a slight decline, which is due to continuous stripping of gas and stars through hydrodynamic drag and tidal torques, respectively (see \cref{fig:cumulmass_sfr} for a comparison between the total stellar mass buildup due to star formation and the actual stellar content reduced by stripping). Some dwarfs exhibit more rapid mass loss after some time, e.g. one from simulation D, where its mass is declining from $\eshort{4}{8}\,\Msun$ to $10^8\,\Msun$ after $t=\SI{1.5}{\giga\year}$. These dwarfs are close to the galaxy cluster center at this point, where they are efficiently shredded by the cluster potential.

\begin{figure*}[ht!]
    \centerline{\includegraphics[width=0.9\textwidth, trim={0 0 0 0},clip]{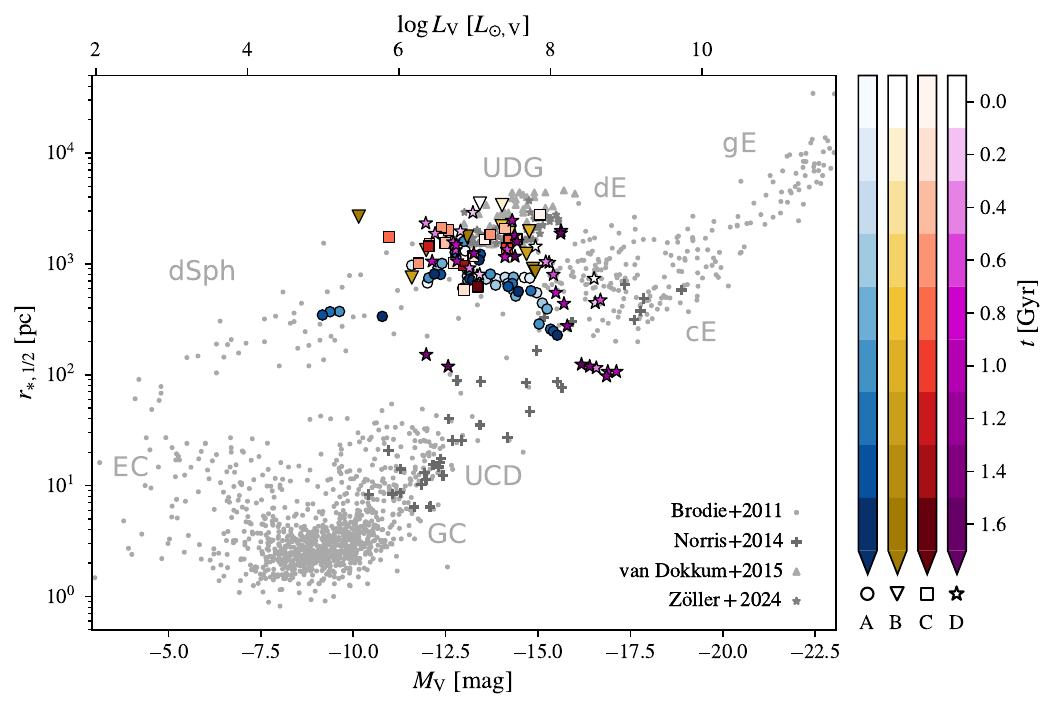}}
    \caption{Stellar half-mass radius $r_{\ast, 1/2}$ vs. absolute (V-band) magnitude $M_V$  of dwarf galaxies. The colored markers show the simulated objects, where the shade indicates the time between $t=0-1.6\, \rm Gyr$ that has passed after the stripping event in the simulations. Gray circles display an observational sample by \citet{Brodie:2011} of giant, compact, and dwarf ellipticals (gE, cE, and dE), dwarf spheroidals (dSph), ultra-compact dwarfs (UCD), as well as globular and extended clusters (GC and EC). The pluses indicate the cE and UCD sample puplished by \citet{Norris:2014}. The gray triangles and stars refer to ultra-diffuse galaxies (UDGs) in the Coma cluster by \citet{vanDokkum:2015} and \citet{Zoeller:2024}, respectively.}
    \label{fig:moneyplot}
\end{figure*}

While the total mass stays relatively constant, the gas content of the dwarfs in simulation A and D is continuously converted into stars, as becomes apparent when comparing the middle and right panel in \cref{fig:massevol}. When comparing A with D, the stellar and gas mass of these dwarfs increases and decreases at similar paces, respectively. It is worth pointing out two outliers from this behavior in simulation D, where the dwarf galaxies' gas mass drops dramatically around $t=\SI{1.5}{\giga\year}$. These are two very interesting cases, where the gas body was suddenly stripped from the stellar core by the ICM -- in line with the very effective gas stripping from the merger remnant that is visible in \cref{fig:zoomoutcomp}. This happens, in fact, to the two most massive objects in this sample. One of them manages to steady itself despite loosing a significant portion of mass that builds its gravitational potential, which is necessary in order to stabilize against tidal forces in the cluster. It continues to persist for the following Gyr, maintaining a stellar (total) mass close to $10^{9}\, \Msun$. The remaining mass of the other dwarf, on the other hand, is not sufficient for such self-shielding, and it begins to develop stellar streams, progressively decreasing the dwarf's stellar content after peaking with $M_\ast = \eshort{3}{8}\,\Msun$ at $t=\SI{2}{\giga\year}$. In this regard it is indeed curious to see, that the aforementioned stripped gas content does not dissolve in the ICM -- instead these clouds stay confined and continue to move as independent dwarf galaxies as a quiescent gas cloud ($M_\ast=\eshort{1.3}{6}\,\Msun$ and $M_{\rm}=\eshort{6}{7}\,\Msun$ at $t=1.5\,\rm Gyr$) in one case, and new star forming dwarf galaxy ($M_\ast=\eshort{1.5}{7}\,\Msun$ and $M_{\rm}=10^8\,\Msun$ at $t=2\,\rm Gyr$) in the other. These two examples are an engaging demonstration of the interesting details in dwarf galaxy evolution inside galaxy clusters when viscosity is not negligible, smoothing out turbulent modes and thus allowing for low surface brightness galaxies in much smaller mass regimes. We will return to this aspect in the Discussion (\cref{subsec:cc_comp}).

The inviscid simulations B and C, however, display a very different behavior compared to A and D. As established above, the dwarfs in setup C are very unstable and disassemble quickly after being initiated by the galaxy merger. On the other hand, the objects in simulation B also struggle to form stars in the beginning, since they arise from overdensities in the tidal tail that are more diffuse than in cases A and B. The only exception from this general behavior in simulation B is the most massive halo, which kickstarts star formation at the same pace as in simulation A and D. But after a few Gyr, they surprisingly also become star forming after having orbited as almost purely gas-dominated (dark) galaxies. Thus, turbulence in simulation B is strong enough to delay collapse and subsequent star formation, but not powerful enough at the relevant scales to disrupt the gas cloud. This fickle equilibrium is disturbed by the additional stellar winds acting on the dwarf galaxies in simulation C, and finally causing them to break apart after their gas content is expelled. We note that, strictly speaking, the stellar wind prescription used in simulation C has been tuned for more massive galaxies in the Milky Way regime, not for dwarf galaxies. Therefore, these results should not be interpreted such that stellar winds always destroy dwarf galaxies. Rather, these results hold if the star forming episode of a dwarf is violent enough to justify rapid and efficient gas evacuation.

\subsubsection{Emerging dwarf galaxy types}\label{subsubsec:dwarftypes}

Finally, we will analyze what kind of dwarf galaxies are formed in the different simulations. Since the dwarfs' sizes and masses vary drastically, it is beneficial to regard them in the stellar half-mass radius $r_{\ast,1/2}$ vs. luminosity $L_V$ plane, as done in \cref{fig:moneyplot}. Following the same procedure as in \citetalias{ivleva+24}, we are comparing observational samples of dwarf galaxies (gray) with our simulated objects (colored), where the circles show the extensive catalogue by \citet{Brodie:2011} that spans over many different dwarf galaxy types, while the triangles and stars display ultra-diffuse galaxies in the Coma cluster by \citet{vanDokkum:2015} and \citet{Zoeller:2024}, respectively.  The luminosity of the simulated dwarf galaxies was estimated by applying age-dependent mass-to-light ratios provided by \citet{Sextl:2023} for each stellar particle in the simulation. In this figure, we only include objects from the simulations, which host at least five stellar particles at the given time. As in the SFR and mass evolution comparison (\cref{fig:massevol,fig:sfr}), the dwarfs from different simulations are distinguished by different markers and colors (blue circles for A, yellow triangles for B, red squares for C and purple stars for D). Note, however, that we now show the dwarfs' evolution over a course of \SI{1.6}{\giga\year} and in more detail, where subsequently darker shades of the respective color indicate time steps of \SI{200}{\mega\year} for the dwarf population in the given simulation.

As already elaborated \citetalias{ivleva+24}, the population in \mbox{simulation A} covers a wide range of dwarf galaxy types. Starting from an initial stellar size in the order of a kpc, they evolve from diffuse, low-surface brightness dwarfs to compact ellipticals over a time span of a few \SI{100}{\mega\year} (see \cref{fig:massevolwsize} for the dwarfs' mass evolution including their stellar size $r_{\ast,1/2}$). We already stated in \cref{subsubsec:massevol}, that the sample from A and D experience very similar mass evolutions and star formation histories, yet it is validating to see that both setups of high effective viscosity give rise to similar dwarf types, in particular producing dwarf spheroidals (dSph) and compact elliptical dwarfs (cE). Nevertheless, though the qualitative behavior is the same, the different subgrid implementations still lead to subtle differences here. While the gas masses of the objects are similar in setups A and D (see right panel in \cref{fig:massevol}), the dwarf's star-forming gas component is more extended in D, which automatically reflects in their stellar radius, leading to a systematic shift towards larger stellar sizes $r_{\ast,1/2}$ in \cref{fig:moneyplot} for simulation D.

In contrast to that, simulations B and C almost exclusively populate the diffuse dwarf galaxy regime (yellow triangles and red squares). The almost vanishing viscosity acting on the dwarf galaxies here leads to more efficient mixing between the objects' boundary layers and the ICM (c.f. \cref{subsec:generalgasbehavior,subsec:excess}). Similar to \mbox{simulation D}, these dwarfs become quite extended in their stellar size. However, they cannot confine their gas component as efficiently against environmental influences, preventing them to accumulate the dense stellar core, which is characteristic for the samples in A and D after about a Gyr.

Comparing the position of the dwarfs in B, C and D to the location of observed dwarf types in \cref{fig:moneyplot}, it is quite intriguing to see that these simulated objects are now clearly overlapping with UDGs. This conformance is especially persistent for setups B and C, i.e. for the realizations of an inviscid ICM. We discuss further similarities and other possible observational counterparts in \cref{subsec:obsanalogs}.

\section{Discussion}

Our conducted simulation campaign allows a unique analysis of the possible gas cloud behaviors and stability trends inside an ICM, while simultaneously granting the possible evolutionary tracks of baryon-dominated dwarf galaxies inside a cluster environment. We discuss both aspects in the following sections.

\subsection{Comparison to cloud crushing studies} \label{subsec:cc_comp}

As established in the Introduction, investigating the conditions for cold cloud entrainment in a hot surrounding medium is a large and active field of reaseach in its own. A multitude of semi-analytical and simulational works has been conducted so far, in the effort to describe the dynamics and entrainment conditions of cold clouds. These studies consider in varying complexity the plethora of circumstances that may influence the cloud's evolution \citep[e.g. presence of tubulent flows, radiative cooling, magnetic fields etc., for a comprehensive review of this field we refer to][]{GronkeSchneider:2026}. No consesus regarding the stability criteria for cold clouds in hot diffuse media has been found yet. Curiously enough, the majority of these studies has been exclusively performed assuming interstellar and circumgalactic medium-like conditions, leaving the parameter space appropriate to galaxy cluster environments largely unexplored yet. The simulations presented in this work hence provide a novel probe of possible mixing behaviors in ICM environments. In particular, these simulations are inherently self-consistent in the regard that we were not forced to use common simplifications such as e.g. constant/vertically stratified background densities or single cloud injections. Instead, our setup \enquote{automatically} models realistic conditions by injecting cold clouds naturally through fully resolved merger and stripping processes, as well as using a cosmological code with its radiative cooling and star forming prescriptions, that have already been exhaustively tested on their consistency in galaxy formation simulations.

So far, only few studies have considered inflowing gas dynamics (approriate to our setup), rather than outflowing gas clouds, which might require differing frameworks. In this regard, \citet{Tan:2023} have recently explored cold clouds exposed to external gravitational fields. As they pointed out, maintaining cloud stability is even more difficult under such circumstances. This is because an outlowing cloud eventually becomes comoving by the hot wind, quenching destructive shear instabilities, while infalling clouds keep accelerating until they reach their terminal velocity with respect to the surrounding medium. Only when a cloud can withstand instabilities at these given velocities, it will survive. They conclude that the commonly used stability crierion $t_{\rm cool,mix}/t_{\rm cc}<1$ (i.e. the cooling or mixing timescale $t_{\rm cool,mix}$ needs to be smaller than the cloud crushing timescale $t_{\rm cc}$) established in outflowing cloud studies is not valid in the infalling gas regime. Instead they formulate a different survival threshold in case if the cloud is able to accrete hot halo gas onto its core, which is appropriate for high cooling efficiencies of the \enquote{warm} gas in the outer turbulent layers. This mass flux then implies a terminal velocity of 
\begin{equation} \label{eq:terminalvel_tan}
    v_{\rm Tan+2023} = \frac{M}{\dot{M}} g,
\end{equation}
\noindent where $M$, $\dot{M}$ and $g$ are the gas cloud's mass, accretion rate and gravitational acceleration, respectively. \citet{Tan:2023} argue that this value will generally be lower than the drag-induced terminal velocity, decreasing the amount of instabilities a cloud has to endure.

Most dwarf galaxies in our simulations are massive enough to maintain angular momentum with respect to the cluster and have not reached terminal velocities before becoming stellar dominated. However, in all three long-term setups setups A, B and D we find one low-mass dwarf where such a state is reached and they start drifting with near constant speed. These dwarfs display slowly declining gas masses in \cref{fig:massevol}, which may be a convolution of gas depletion due to star formation, as well as potential gas loss or accretion. We have checked for these two candidates the respective contribution by estimating
\begin{equation}
    x = \frac{\Delta M}{\text{SFR} \cdot \Delta t},
\end{equation}
\noindent where $\Delta M$ is the mass loss during a time span of $\Delta t$ ($=\SI{0.5}{\giga\year}$ according to our tracing time steps) and SFR is the dwarf's star formation rate\footnote{We could use that the SFR is relatively constant across several Gyr, c.f. upper panel in \cref{fig:sfr}.}. For the clouds in setups A and B, this value ranges between $x_{\rm A,B}=1-2$, meaning that there is additional gas stripping contributing to the gas mass loss apart from star formation, and \cref{eq:terminalvel_tan} by \citet{Tan:2023} is not applicable. For the dwarf in simulation D, this value is lower around $x_{\rm D}=0.8-1$, indicating that mass loss is mostly balanced by star formation, though minor gas accretion is occurring. If we now interpret this relative mass gain as the accretion rate $\dot{M}$, we would arrive at a mass accretion-limited terminal velocity of $v_{\rm Tan+2023}\approx \SI{4500}{\kilo\meter/\second}$.

In order to compare this to the drag limited terminal velocity, let us consider force balance between gravity and drag, i.e.
\begin{equation} \label{eq:drageq}
    Mg - \frac{1}{2}\rho_{\rm ICM} C A v^2 - \Pi_{\rm eff} = 0,
\end{equation}
\noindent where $\rho_{\rm ICM}$ is the density of the  surrounding hot medium, and $C$ the geometric drag coefficient, while $A$ and $v$ are the cross section and velocity of the cloud. The term $\Pi_{\rm eff}$ encapsulates complementary drag forces apart from the classical geometric drag. If $\Pi_{\rm eff}=0$, the terminal drag velocity will be
\begin{equation}
    v_{\rm drag} = \sqrt{\frac{8\chi R g}{3C}},
\end{equation}
\noindent where $R$ is the radius of the spherical cloud and $\chi=\rho_{\rm cold} / \rho_{\rm ICM}$ its density contrast with respect to the ICM. The value of $C$ strongly depends on the Reynolds number $Re$ of the flow and is equal to $C=1-20$ for $Re = 100 - 1$ in the case of a fluid around a sphere \citep[c.f. Fig. 34 by][]{LandauLifschitzHydro}. For reference, we estimated the typical Reynolds number $Re = vR\rho/\eta$ for the mentioned dwarf galaxy in our simulation, using the Spitzer value as the shear viscosity coefficient $\eta$, where the ambient density is $\rho_{\rm ICM}=\eshort{4}{-28}\,\rm g/cm^3$ and the initial velocity of the cloud is in the order of the merger parent drift velocity inside the cluster $v\sim1000\,\rm km/s$. For typical temperatures between $T=1-\eshort{2}{7}\,\rm K$, the minimal Reynolds number in simulation D is therefore $Re=6-1$, while higher $Re$ are applicable to dwarfs in less viscous realizations (simulation B+C). At the dwarf's distance from the cluster ($\sim$\SI{510}{\kilo\parsec}), we computed the gravitational acceleration by the cluster to be $g=\eshort{3}{-9}\, \rm cm/s^2$, while the density contrast is $\chi=2500$ with a cloud radius of $R\sim \rm kpc$. Hence, the terminal drag velocity will be $v_{\rm drag}=560-2500\,\rm km/s$ for $Re = 1-100$. We measure the actual final drift velocity of the dwarf galaxy to be $v_{d} \approx \SI{150}{\kilo\meter/\second}$. Hence, it is drifting significantly slower than both the mass-accretion limited velocity $v_{\rm Tan+2023}$ and terminal drag velocity estimate $v_{\rm drag}$ -- even for extremely viscous flows with $Re\sim1$. The former issue is no surprise, considering the low overall mass accretion rate of the dwarf, which would translate into a very high mass doubling time scale of $t_{\rm grow}=M/\dot{M}\sim\SI{5}{\giga\year}$. Thus, this effect is acting much slower than any other relevant process such as e.g. the variability of the background density and temperature, or of the cloud's internal properties like its shape and SFR.

So, if mass accretion onto the cloud is not driving the lower final drift velocity of the cloud, what else is deccelerating it and causing it to survive? Neglecting further drag terms $\Pi_{\rm eff}$ is valid only in the case of flows around rigid bodies, captured by the second term in \cref{eq:drageq}. Mass accretion onto the cold cloud can be interpreted as such an effective drag term $\Pi_{\rm eff}$, since it is causing momentum transfer from the cold to the hot medium (ICM gas initially at rest starts comoving with the cloud). But apart from that, the rigid body approximation is certainly not true for gas clouds, which are susceptible to deformation. Addionally, shear instabilities at the envelope can increase the effective cross section of the cloud, translating into significantly larger drag coefficients. We point out that we measured a similar final drift velocity for the viscous (A+D) and invscid (B) setup, indicating that the gas compression caused by the dwarf's motion is strong enough to trigger the artificial bulk viscosity scheme in simulation B, effectively causing it to behave similar as in the fully viscous realizations. We did not include magnetic fields in our simulations, but they have already been shown in multiple studies to balance the cloud against instabilities if the field strengths are large enough \citep[e.g.,][]{Hidalgo-Pineda:2024,Kaul:2025}. Hence, our results can be interpreted as the \enquote{worst case} outcome for cloud stability.

In summary we conclude, that giant gas clouds and gas-rich dwarf galaxies are remarkably stable inside the ICM due to relatively low terminal velocities in the cluster potential, induced by hydrodynamic drag. They can easily survive for several Gyr if they have initial masses of at least $M_{\rm gas}\gtrsim10^7\,\Msun$. Our tested values for effective viscosity cover the whole physically feasable range from practically inviscid (setup B) to full Spitzer value (setup D). Therefore, our stability prediction holds for any galaxy cluster no matter its effective viscosity, though we note that the lower mass limit for stable gas clouds in invscid ICM is higher by about a factor of 2 (\cref{fig:massevol}). We will test whether our inferred mass threshold is due to resolution limitations in future work.

\subsection{Analogs to unusual dwarf galaxies in observations}\label{subsec:obsanalogs}

According to basic classification in terms of their stellar size and luminosity, long-lived tidal dwarf galaxies (TDGs) evolve into a large variety of dwarf galaxy types across all tested subgrid implementaitons for viscosity and feedback (\cref{fig:moneyplot}). Meanwhile, they all have in common an elevated star formation rate (SFR) with respect to the star forming main sequence. Our setup is intrinsically independent of reshift, since we have performed the simulation in an idealized setup. Hence our predictions can be applied thoughout cosmic time, as long as the circumstances are realistic for the respective epoch. These are in particular (i) gas-rich merger events and (ii) the presence of a massive enough galaxy group or cluster, which can exert ram pressure. While the former condition is more often met in the early Universe, the latter is not satisfied when going to too high redshifts, since structure formation there has not progressed far enough to form galaxy clusters yet. In this regard we note that, while having systematically lower SFRs than the high redshift dwarfs in \cref{fig:sfr}, there is still an overlap between our simulated sample and these observed galaxies, indicating the universality in redshift of the formation channel tested in this work.

We stress that the simulated dwarf galaxies here are surely not counterparts to normal dwarf galaxies. On the other hand, the examined formation scenario recovers key properties of some peculiar dwarf galaxies found in low-surface brightness surveys and as blind radio searches. Many of these observations point towards potentially important cluster influences through ram pressure and tidal torques. In this work, we conducted a first of its kind simulation campaign, where we self-consistently tested the influence of clusters on the dynamics of dwarf galaxies in high spatial and time resolution. In the following we will discuss possible analogs from our simulations to such peculiar observed galaxy types, namely blue candidates (BCs), dark galaxies and ultra-diffuse galaxies (UDGs).

\subsubsection{Blue candidates (\enquote{blue blobs})}

BCs are characterized by their clumpy morphology with very small stellar masses (typically between $10^4 - 10^5\,\Msun$) and they seem to appear in associations inside the galaxy cluster, while being separated from any other larger galaxy by at least a few \SI{100}{\kilo\parsec} \citep{Dey:2025}. Their elevated SFR and metallicity with respect to the star forming main sequence and mass-metallicity relation for dwarf galaxies indicates that they are not normal galaxies, i.e. they do not seem to have formed early on in the Universe according to hierarchical structure formation. Instead, two formation scenarios have been proposed by \citet{Jones:2022}, which interpret BCs as either ram pressure-stripped gas from larger galaxies or long-lived TDGs. Considering the former possibility, many wind tunnel simulations have demonstrated that star forming clumps can form in jellyfish-like gas filaments behind galaxies, which in principle could later on detach and fashion small clusters of blue dwarfs. However, \citet{Jones:2022} points out that the objects produced by this channel display gas masses that are at least an order of magnitude lower than the HI content measured in BCs \citep[usually $\geq10^7\,\Msun$,][]{Dey:2025}. The apparent isolation of BCs inside the cluster with respect to larger galaxies would be difficult to reconcile with as well.

Long-lived TDGs, on the other hand, should have masses of at least $10^8\,\Msun$ according to the merger simulations by \citet{BournaudDuc:2006}, thus being too massive compared to observed BCs. Apart from that, the large separation between BCs and possible parent galaxies could not be recovered with typical rotational velocities of the parent (taken as the ejection velocity of stripped TDGs) and the stellar age of BCs. However, we have recently demonstrated in \citetalias{ivleva+24} that a cluster environment drastically increases the TDG production rate and reduces the mass end for long-lived TDGs compared to the isolated setup by \citet{BournaudDuc:2006} due to ram pressure-assisted stripping from the parent. Additionally, the relative velocity between the parent galaxy and stripped TDGs is better descibed by the drift velocity of the parent inside the cluster, since the gas-dominated dwarfs are quickly decelerated by fluid drag. Therefore, they can easily reach distances of several \SI{100}{\kilo\parsec} with respect to the merger remnant in $\sim$\SI{100}{\mega\year} (c.f. Figure 8 in \citetalias{ivleva+24}), while remaining in loose associations with each other, agreeing with the apparent isolation of BCs. Though they are below the resolution limit in our simulations, our sample produces similarly elevated SFR with respect to the star forming main sequence, with remarkable agreement between the extrapolation of the simulated SFRs towards lower stellar masses and observed BCs (see \cref{subsubsec:sfr}). Moreover, some dwarf galaxies in our sample -- in particular from setup B -- display similarly large gas to stellar mass ratios between $M_{\rm gas}/M_\ast \approx 100-160$, agreeing well with BC estimates, e.g. $M_{\rm gas}/M_\ast \approx 150$ for \mbox{SECCOI} \citep{Jones:2022}. Their location inside the phase space diagram (line of sight velocity vs. projected distance to cluster center) in intermediate infall time regions despite their young stellar age is compatible with our simulated objects as well (\mbox{Ivleva et al.} in prep.). We will push the simulation towards higher resolution in future work, finalizing the confirmation of the hypothesis by \citet{Jones:2022}, according to which BCs can be interpreted as long-lived TDGs.

\subsubsection{(Almost) dark galaxies}

There might be an interesting overlap between BCs and dark galaxy candidates, i.e. objects that seem to be entirely gas dominated without detected counterpart in optical or UV. Some objects previously classified as dark galaxies turned out to actually host a small stellar component, that has not been visible with the sensitivity at hand before \citep[e.g.,][]{Adams:2015,Bellazzini:2015,Cannon:2015,Jones:2022b,Jones:2022}. Indeed, we also find several large gas clouds in our simulations, which display a very low amount of stars early-on after the stripping event. In particular, \citet{Gray:2023} has recently explored almost-dark galaxies from the ALFALFA survey and concluded that two candidates are likely TDGs based on their low dark matter content. These galaxies are not in close association to any larger galaxy, as it is the case in our simulated objects. Also their inferred stellar and gas masses fit well with our simulated dwarf sample, particularly from setup B. This combination supports that our tested formation channel indeed contributes to the variety of dwarf galaxies in the local Universe.

We note that the evolutionary pathway examined in this work requires a stripping agent in the form of a galaxy cluster's ICM or a massive galaxy group environment with assembled hot halo, which can remove gas from infalling galaxies. Systematic searches across the sky for dark galaxy candidates in blind HI surveys are still underway. Thus, it cannot be finally concluded yet whether dark galaxies appear in specific environments, though \citet{Kwon:2025} report a slight tendency towards underdense regions (c.f. their \mbox{figure 7}). As such, our tested formation pathway should only be interpreted as one possible channel to form dark galaxies in overdense environments, while clearly additional mechanisms must contribute in void regions. Upcoming data releases might clarify tentative environmental trends.

\subsubsection{Dark matter-deficient UDGs}

We have demonstrated in \cref{subsubsec:dwarftypes} that all of our setups produced objects resembling UDGs according to the standard stellar size and surface brightness criterion. While being currently quiescent, recent spectroscopic follow-up observations on UDGs have inferred bursty star formation histories through metallicity abundances, where their stellar masses seem to have been formed just in a single episode in some cases \citep[e.g.,][]{Ferre-Mateu:2023,Gannon:2026}. Inferring their respective peak SFR is difficult due to lacking time resolution in the reconstructed histories. Yet, there are also several currently star-forming, HI-rich UDGs observations published, that report SFRs comparable to main sequence dwarfs \citep{Kado-Fong:2022}. When comparing them to the simulated dwarfs, we see that our objects tend to have higher SFRs, however there is a noticeable overlap with the most star forming instances of the observed UDG sample and \mbox{setup D} (see lower panel in \cref{fig:sfr}). Considering that this simulation also recovers many UDG-like dwarfs in \cref{fig:moneyplot}, we propose that a peculiar subsample of such UDGs could be long-lived TDGs, if they are found in cluster environments and display a dark matter deficiency.

We note that the inviscid \mbox{simulation B} also developed a relatively large number of UDG-like dwarfs, which persevere in such a state for a longer time, whereas most galaxies from \mbox{setup D} turn into compact dwarfs during the first few Gyr of their evolution (\cref{fig:moneyplot}). However, these dwarfs from \mbox{setup B} also displayed higher SFRs than observed in currently star forming UDGs (\cref{fig:sfr}). Future spectroscopic studies on quiescent UDGs might reveal clearer time evolutions, which will allow inferring their peak SFRs in the past and test agreement with our predicted SFRs for such dwarfs. Additionally, it should be pointed out that the coverage of most currently available studies on UDGs in galaxy clusters do not reach beyond \mbox{$\sim$$0.5\,\rm R_{\rm vir}$}, while a major prediction of the tested channel is that gas-rich, dark-matter deficient UDGs can be expected in the cluster outskirts between $0.5-1\,\rm R_{\rm vir}$.

\section{Summary and Conclusion}\label{sec:sumcon}

We have conducted four high resolution galaxy merger simulations inside a realistic cluster environment, where we have varied the stellar feedback and viscosity treatments across the physically feasable parameter space. Thus our simulations bracket the extremes of possible circumstances, where we modelled both a practically inviscid and very viscous ICM, which has major consequences for multi-phase mixing applicable to cold gas clouds and baryon-dominated dwarf galaxies inside cluster environments. Different stellar feedback implementations allowed us to investigate the impact of low vs. high effective feedback on dwarf galaxies.

We confirm our previous result from \citetalias{ivleva+24} and verify that stripping can efficiently remove TDGs from the merger remnant in all realizations. The resulting gas clouds and dwarf galaxies are remarkably stable across all simulations, except when strong stellar feedback is acting. With moderate stellar feedback, all setups produce dwarfs with similar total masses ranging between $M_{\rm tot}=10^7-10^9\,\Msun$. Low viscosities generally lead to more diffuse and gas-dominated objects, while dwarfs in high viscosity ICM display higher stellar to gas mass ratios, turning into compact, stellar dominated dwarf galaxies after a few Gyr.

The smallest gas clouds in the viscous representations reached terminal drift velocities inside the cluster of $v_{d}\sim100\,\rm km/s$. This is significantly slower than values motivated by geometric drag even for extremely viscous flows and thus quenches disruptive surface instabilities. The minimal Reynolds number that is applicable for these gas clouds is $Re\sim1$.

All dwarf galaxies display elevated SFRs with respect to the main sequence of dwarfs inside clusters. For the viscous realizations, the SFRs stay relatively stable across several Gyr around $0.01-0.1\,\Msun/\rm yr$. Clouds in the inviscid regime are more quiescent in the beginning, but display rising SFRs after $\sim$\SI{1}{\giga\year}, before converging to the same value as in the viscous scenarios.

Based on their properties in SFR, stellar/gas mass, stellar size and luminosity, we propose that a subset of peculiar galaxies observed inside galaxy clusters -- namely blue candidates (BCs), dark galaxies and dark matter deficient ultra-diffuse galaxies (UDGs) -- are stripped, long-lived TDGs. The elevated SFR of our simulated galaxies is also compatible with observed dwarfs at high redshift ($3<z<5$), indicating that this formation channel has been active at early cosmic times as well.

\begin{acknowledgements}
We thank Ildar Khabibullin, Michael G. Jones and Marco Monaci for helpful suggestions improving the quality of the manuscript. AI and TM acknowledge support by the COMPLEX project from the European Research Council (ERC) under the European Union’s Horizon 2020 research and innovation program grant agreement ERC-2019-AdG 882679. AI and KD  acknowledge support by the DFG project Nr. 516355818. DF thanks the ARC for financial support via DP250101673.

This research was supported by the Excellence Cluster ORIGINS, funded by the Deutsche Forschungsgemeinschaft under Germany's Excellence Strategy -- EXC-2094-390783311.

AI, RSR and DAF acknowledge support from the German exchange program DAAD-PPP under the Project Number 57750566.

The following software was used for this work: \textsc{Julia} \citep{bezanson+17:julia}, \textsc{Splotch} \citep{Dolag:2008}, \textsc{Matplotlib} \citep{Hunter:2007}.
\end{acknowledgements}

\bibliographystyle{style/aa_url}
\bibliography{bib}

\begin{appendix}

\section{Cumulative stellar mass growth}

\begin{figure*}[ht!]
    \centerline{\includegraphics[width=\textwidth, trim={0 0 0 0}]{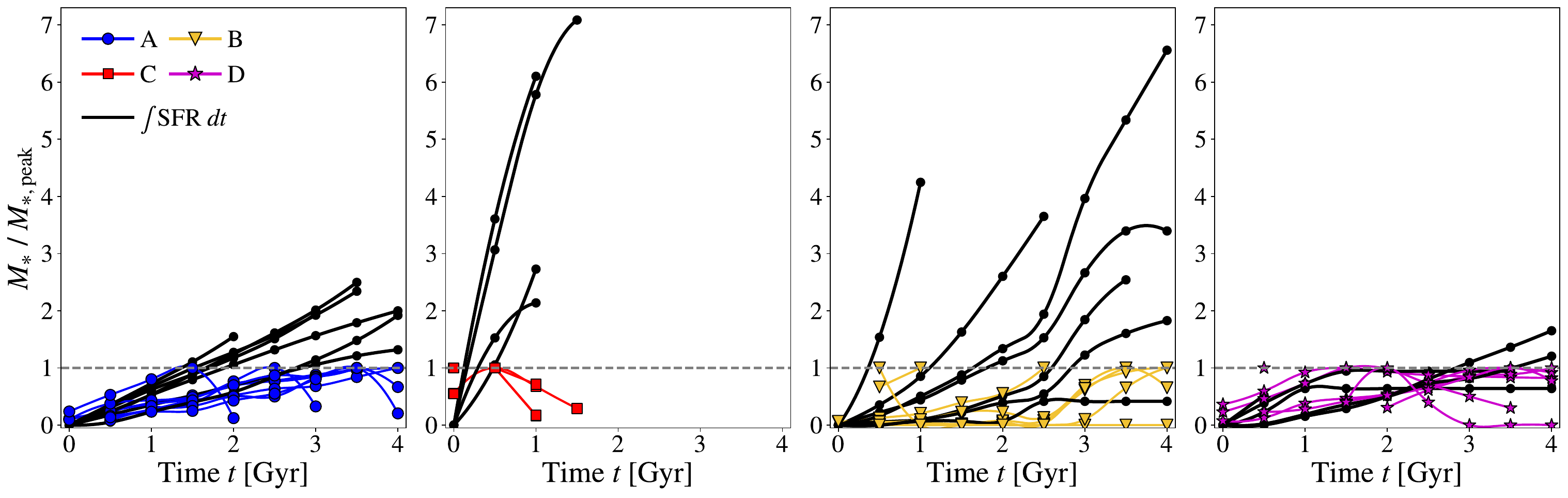}}
    \caption{Cumulative stellar mass over time, normalized by the dwarfs' peak stellar mass. The colored lines show the actual evolution, while the black lines indicate the galaxies' stellar body expected from its star formation rate. It becomes apparent that tidal stripping is far more efficient in the inviscid setups B and C.}
    \label{fig:cumulmass_sfr}
\end{figure*}

\cref{fig:cumulmass_sfr} presents the cumulative stellar mass growth of the simulated dwarfs separately for the four samples (A-D) in each panel with the colored lines. Each dwarf is normalized on its peak mass, i.e. the maximum stellar mass it ever acquired. As described in \cref{subsubsec:sfr}, the evolution between the tracing times (shown by the markers) is indicated by a fitted spline. The black lines show the stellar mass expected from the dwarf's star formation rate. While the viscous setups A and D lose at most twice their total stellar body due to stripping, the inviscid realizations B and C display much more stripping, which leads to pronounced tidal streams.

\section{Mass evolution with stellar size}

\begin{figure*}[ht!]
    \centerline{\includegraphics[width=\textwidth, trim={0 0 0 0}]{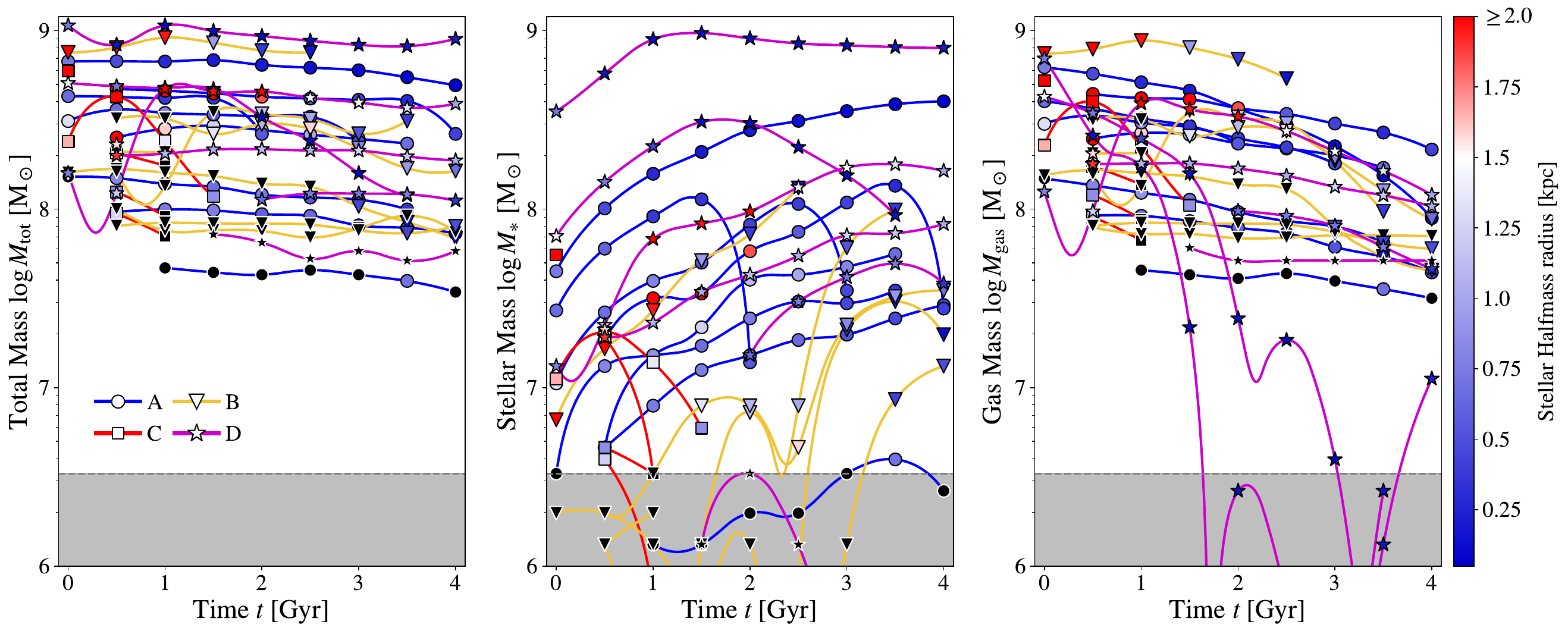}}
    \caption{Mass evolution correlated with stellar size. We show the same as in \cref{fig:massevol}, but here the markers at the tracing times are colored according to the galaxy's stellar half-mass radius. Black color means that the dwarf has less than six stellar particles.}
    \label{fig:massevolwsize}
\end{figure*}

In \cref{fig:massevolwsize} we show the same as in \cref{fig:massevol}, but include information about the dwarf's stellar half-mass radius here, indicated by the color of the marker at a given tracing time. As time passes, the dwarfs from setup A and D gradually evolve from diffuse to compact galaxies. The most massive object from setup B displays the same behavior. The other dwarfs from this sample, however, maintain mostly gas-dominated with low-mass, compact stellar components. Simulation C destroys its dwarfs efficiently due to high effective stellar feedback after \SI{1.5}{\giga\year}.

\section{Dwarf data}

The four subtables in \cref{tab:dwarfprops} list the properties of the identified dwarf galaxies of each of the four simulations (A-D), respectively. The properties are gas mass $M_g$, stellar mass $M_\ast$, stellar half-mass radius $r_{\ast,1/2}$, star formation rate SFR and V-band luminosity $L_V$, and are given for the full tracing period of $4\,\rm Gyr$ ($1.5\,\rm Gyr$ for simulation C) with a time step of $0.5\,\rm Gyr$. An entry of \enquote{--} means the dwarf is not present yet or anymore at the given timestep, while \enquote{n.a.} (i.e. not applicable) in the columns for $r_{\ast,1/2}$ and SFR denotes cases where the object has less than five stellar particles ($\log M_\ast / \Msun < 6.52$) or a vanishing star formation rate, respectively.

We only list the first three dwarf galaxies from each setup here, the full tables can be downloaded in the online version of the article. The simulations A, B, C and D produced eight, seven, five and seven dwarfs, respectively. 

\begin{table*}
\setlength{\tabcolsep}{3pt}
\centering
\renewcommand{\arraystretch}{1.25}
\subfloat[Simulation A]{
\resizebox{\textwidth}{!}{%
\begin{tabular}{c | ccccc|ccccc|ccccc}
\hline\hline
ID & \multicolumn{5}{c|}{1} 
& \multicolumn{5}{c|}{2} 
& \multicolumn{5}{c}{\hfill3\hfill$\boldsymbol{\rightarrow}$} \\
\hline
$t$ 
& $\log(M_{\mathrm{g}})$ & $\log(M_{*})$ & $\log(r_{*})$ & $\log(\mathrm{SFR})$ & $\log(L_V)$
& $\log(M_{\mathrm{g}})$ & $\log(M_{*})$ & $\log(r_{*})$ & $\log(\mathrm{SFR})$ & $\log(L_V)$
& $\log(M_{\mathrm{g}})$ & $\log(M_{*})$ & $\log(r_{*})$ & $\log(\mathrm{SFR})$ & $\log(L_V)$ \\
 Gyr & $[\mathrm{M}_\odot]$ & $[\mathrm{M}_\odot]$ & $[\mathrm{kpc}]$ & $[\mathrm{M}_\odot\,\mathrm{yr}^{-1}]$ & $[\mathrm{L}_{V,\odot}]$
& $[\mathrm{M}_\odot]$ & $[\mathrm{M}_\odot]$ & $[\mathrm{kpc}]$ & $[\mathrm{M}_\odot\,\mathrm{yr}^{-1}]$ & $[\mathrm{L}_{V,\odot}]$
& $[\mathrm{M}_\odot]$ & $[\mathrm{M}_\odot]$ & $[\mathrm{kpc}]$ & $[\mathrm{M}_\odot\,\mathrm{yr}^{-1}]$ & $[\mathrm{L}_{V,\odot}]$ \\
\hline
0.0 & -- & -- & -- & -- & -- & 8.79 & 7.65 & -0.12 & -1.02 & 7.78 & 8.48 & 7.02 & 0.12 & -1.44 & 7.14 \\
0.5 & 8.39 & 6.66 & 0.30 & -1.61 & 6.71 & 8.76 & 8.00 & -0.32 & -0.84 & 8.02 & 8.52 & 7.42 & -0.17 & -1.18 & 7.38 \\
1.0 & 8.43 & 7.18 & 0.20 & -1.45 & 7.13 & 8.71 & 8.20 & -0.54 & -0.85 & 7.94 & 8.49 & 7.60 & -0.09 & -1.17 & 7.42 \\
1.5 & 8.42 & 7.34 & 0.10 & -1.41 & 7.45 & 8.66 & 8.32 & -0.69 & -0.87 & 8.11 & 8.46 & 7.70 & -0.26 & -1.25 & 7.66 \\
2.0 & 8.37 & 7.60 & 0.05 & -1.40 & 7.53 & 8.56 & 8.44 & -0.70 & -0.85 & 8.14 & 8.40 & 7.91 & -0.38 & -1.21 & 7.59 \\
2.5 & 8.34 & 7.63 & 0.00 & -1.38 & 7.32 & 8.47 & 8.49 & -0.83 & -0.83 & 8.13 & 8.35 & 8.03 & -0.40 & -1.14 & 7.68 \\
3.0 & 8.30 & 7.68 & -0.09 & -1.35 & 7.31 & 8.35 & 8.55 & -0.95 & -0.85 & 8.15 & 8.26 & 7.54 & -0.33 & -1.03 & 7.34 \\
3.5 & 8.23 & 7.75 & -0.16 & -1.26 & 7.66 & 8.19 & 8.59 & -1.10 & -0.91 & 8.13 & -- & -- & -- & -- & -- \\
4.0 & -- & -- & -- & -- & -- & 7.95 & 8.60 & -1.23 & -1.07 & 8.08 & -- & -- & -- & -- & -- \\
\hline
\end{tabular}
}
}
\\
\subfloat[Simulation B]{
\resizebox{\textwidth}{!}{%
\begin{tabular}{c | ccccc|ccccc|ccccc}
\hline\hline
ID & \multicolumn{5}{c|}{1} 
& \multicolumn{5}{c|}{2} 
& \multicolumn{5}{c}{\hfill3\hfill$\boldsymbol{\rightarrow}$} \\
\hline
$t$ 
& $\log(M_{\mathrm{g}})$ & $\log(M_{*})$ & $\log(r_{*})$ & $\log(\mathrm{SFR})$ & $\log(L_V)$
& $\log(M_{\mathrm{g}})$ & $\log(M_{*})$ & $\log(r_{*})$ & $\log(\mathrm{SFR})$ & $\log(L_V)$
& $\log(M_{\mathrm{g}})$ & $\log(M_{*})$ & $\log(r_{*})$ & $\log(\mathrm{SFR})$ & $\log(L_V)$ \\
 Gyr & $[\mathrm{M}_\odot]$ & $[\mathrm{M}_\odot]$ & $[\mathrm{kpc}]$ & $[\mathrm{M}_\odot\,\mathrm{yr}^{-1}]$ & $[\mathrm{L}_{V,\odot}]$
& $[\mathrm{M}_\odot]$ & $[\mathrm{M}_\odot]$ & $[\mathrm{kpc}]$ & $[\mathrm{M}_\odot\,\mathrm{yr}^{-1}]$ & $[\mathrm{L}_{V,\odot}]$
& $[\mathrm{M}_\odot] $ & $[\mathrm{M}_\odot]$ & $[\mathrm{kpc}]$ & $[\mathrm{M}_\odot\,\mathrm{yr}^{-1}]$ & $[\mathrm{L}_{V,\odot}]$ \\
\hline
0.0 & 8.87 & 6.82 & 0.54 & -1.30 & 7.30 & 8.19 & 6.30 & n.a. & -2.60 & 6.60 & -- & -- & -- & -- & -- \\
0.5 & 8.90 & 7.22 & 0.36 & -1.03 & 7.43 & 8.22 & 6.30 & n.a. & -1.87 & 6.64 & 8.20 & 6.12 & n.a. & -1.96 & 6.25 \\
1.0 & 8.94 & 7.43 & 0.29 & -0.88 & 7.84 & 8.20 & 6.12 & n.a. & -1.87 & 6.47 & 8.13 & n.a. & n.a. & -2.08 & n.a. \\
1.5 & 8.91 & 7.71 & -0.03 & -0.69 & 7.96 & 8.18 & 5.82 & n.a. & -1.68 & 6.31 & 8.54 & 6.12 & n.a. & -1.48 & 6.39 \\
2.0 & 8.84 & 7.86 & -0.43 & -0.59 & 7.85 & 8.14 & 6.12 & n.a. & -1.69 & 6.45 & 8.52 & 6.86 & 0.06 & -1.32 & 7.44 \\
2.5 & 8.73 & 8.12 & -0.80 & -0.56 & 8.15 & 8.11 & 5.82 & n.a. & -1.61 & 6.22 & 8.48 & 6.90 & 0.00 & -1.17 & 7.09 \\
3.0 & -- & -- & -- & -- & -- & 7.90 & 7.32 & -0.36 & -1.16 & 7.50 & 8.27 & 7.78 & -0.27 & -0.71 & 7.95 \\
3.5 & -- & -- & -- & -- & -- & 7.73 & 7.48 & -0.77 & -1.35 & 7.30 & 7.99 & 8.00 & -0.41 & -0.86 & 7.95 \\
4.0 & -- & -- & -- & -- & -- & 7.66 & 7.30 & -0.95 & n.a. & 6.88 & -- & -- & -- & -- & -- \\
\hline
\end{tabular}
}
}
\\
\subfloat[Simulation C]{
\resizebox{\textwidth}{!}{%
\begin{tabular}{c | ccccc|ccccc|ccccc}
\hline\hline
ID & \multicolumn{5}{c|}{1} 
& \multicolumn{5}{c|}{2} 
& \multicolumn{5}{c}{\hfill3\hfill$\boldsymbol{\rightarrow}$} \\
\hline
$t$ 
& $\log(M_{\mathrm{g}})$ & $\log(M_{*})$ & $\log(r_{*})$ & $\log(\mathrm{SFR})$ & $\log(L_V)$
& $\log(M_{\mathrm{g}})$ & $\log(M_{*})$ & $\log(r_{*})$ & $\log(\mathrm{SFR})$ & $\log(L_V)$
& $\log(M_{\mathrm{g}})$ & $\log(M_{*})$ & $\log(r_{*})$ & $\log(\mathrm{SFR})$ & $\log(L_V)$ \\
 Gyr & $[\mathrm{M}_\odot]$ & $[\mathrm{M}_\odot]$ & $[\mathrm{kpc}]$ &$ [\mathrm{M}_\odot\,\mathrm{yr}^{-1}]$ & $[\mathrm{L}_{V,\odot}]$
& $[\mathrm{M}_\odot]$ & $[\mathrm{M}_\odot]$ & $[\mathrm{kpc}]$ & $[\mathrm{M}_\odot\,\mathrm{yr}^{-1}]$ & $[\mathrm{L}_{V,\odot}]$
& $[\mathrm{M}_\odot]$ & $[\mathrm{M}_\odot]$ & $[\mathrm{kpc}]$ & $[\mathrm{M}_\odot\,\mathrm{yr}^{-1}]$ & $[\mathrm{L}_{V,\odot}]$ \\
\hline
0.0 & 8.36 & 7.05 & 0.22 & -1.61 & 7.36 & 8.72 & 7.74 & 0.44 & -0.83 & 7.95 & -- & -- & -- & -- & -- \\
0.5 & 8.60 & 7.31 & 0.56 & -0.90 & 7.44 & -- & -- & -- & -- & -- & 7.96 & 6.60 & 0.09 & -2.08 & 6.68 \\
1.0 & 8.35 & 7.14 & 0.14 & -0.95 & 7.60 & -- & -- & -- & -- & -- & 7.82 & 5.82 & n.a. & -1.88 & 5.38 \\
1.5 & 8.02 & 6.77 & -0.06 & -1.27 & 7.16 & -- & -- & -- & -- & -- & -- & -- & -- & -- & -- \\
\hline
\end{tabular}
}
}
\\
\subfloat[Simulation D]{
\resizebox{\textwidth}{!}{%
\begin{tabular}{c | ccccc|ccccc|ccccc}
\hline\hline
ID & \multicolumn{5}{c|}{1} 
& \multicolumn{5}{c|}{2} 
& \multicolumn{5}{c}{\hfill3\hfill$\boldsymbol{\rightarrow}$} \\
\hline
$t$ 
& $\log(M_{\mathrm{g}})$ & $\log(M_{*})$ & $\log(r_{*})$ & $\log(\mathrm{SFR})$ & $\log(L_V)$
& $\log(M_{\mathrm{g}})$ & $\log(M_{*})$ & $\log(r_{*})$ & $\log(\mathrm{SFR})$ & $\log(L_V)$
& $\log(M_{\mathrm{g}})$ & $\log(M_{*})$ & $\log(r_{*})$ & $\log(\mathrm{SFR})$ & $\log(L_V)$ \\
 Gyr & $[\mathrm{M}_\odot]$ & $[\mathrm{M}_\odot]$ & $[\mathrm{kpc}]$ & $[\mathrm{M}_\odot\,\mathrm{yr}^{-1}]$ & $[\mathrm{L}_{V,\odot}]$
& $[\mathrm{M}_\odot]$ & $[\mathrm{M}_\odot]$ & $[\mathrm{kpc}]$ & $[\mathrm{M}_\odot\,\mathrm{yr}^{-1}]$ & $[\mathrm{L}_{V,\odot}]$
& $[\mathrm{M}_\odot]$ & $[\mathrm{M}_\odot]$ & $[\mathrm{kpc}]$ & $[\mathrm{M}_\odot\,\mathrm{yr}^{-1}]$ & $[\mathrm{L}_{V,\odot}]$ \\
\hline
0.0 & 8.84 & 8.55 & -0.13 & -0.20 & 8.54 & 8.63 & 7.85 & 0.16 & -0.71 & 7.90 & 8.10 & 7.12 & -0.11 & -1.22 & 7.27 \\
0.5 & 8.41 & 8.76 & -0.80 & -0.35 & 8.55 & 8.54 & 8.15 & -0.14 & -0.63 & 8.10 & 7.99 & 7.35 & 0.09 & -2.96 & 7.21 \\
1.0 & 8.24 & 8.95 & -0.97 & -0.10 & 8.78 & 8.40 & 8.35 & -0.36 & -0.68 & 8.21 & 8.59 & 7.83 & 0.39 & -1.27 & 7.65 \\
1.5 & 7.34 & 8.99 & -0.96 & n.a. & 8.44 & 8.20 & 8.49 & -0.91 & -0.87 & 8.15 & 8.56 & 7.92 & 0.30 & -1.32 & 7.59 \\
2.0 & 6.42 & 8.96 & -0.90 & n.a. & 8.27 & 7.39 & 8.48 & -0.98 & n.a. & 7.92 & 8.52 & 7.99 & 0.33 & -1.12 & 7.68 \\
2.5 & 5.82 & 8.93 & -0.89 & n.a. & 8.13 & 7.27 & 8.35 & -0.95 & n.a. & 7.64 & 8.44 & 8.12 & 0.16 & -0.97 & 7.89 \\
3.0 & 5.82 & 8.92 & -0.80 & n.a. & 8.04 & 6.60 & 8.19 & -1.04 & n.a. & 7.38 & 8.32 & 8.23 & 0.17 & -0.98 & 7.95 \\
3.5 & 6.12 & 8.91 & -0.74 & n.a. & 7.97 & 6.42 & 7.97 & -0.82 & n.a. & 7.08 & 8.21 & 8.25 & 0.13 & -1.02 & 7.93 \\
4.0 & 7.05 & 8.90 & -0.67 & n.a. & 7.92 & -- & -- & -- & -- & -- & 8.08 & 8.21 & -0.00 & -0.99 & 7.95 \\
\hline
\end{tabular}
}
}
\caption{Evolution of galaxy properties from all four simulations A-D (c.f. \cref{tab:simtable}). Only the first three dwarfs from each setup are shown (the continuation of the table is indicated by \enquote{$\boldsymbol{\rightarrow}$}), the full tables are available online.}
\label{tab:dwarfprops}
\end{table*}

\end{appendix}

\end{document}